%% file: ms.tex
\documentclass{aa}
\newcommand{\sftw}[1]{\texttt{#1}}
\newcommand{\sgn}[1]{\mathrm{sgn}(#1)}

\usepackage{color}
\usepackage[normalem]{ulem}
\usepackage{amsmath,amssymb}	
\usepackage{textcomp}
\usepackage{booktabs}
\usepackage{xcolor}
\usepackage{nicefrac}
\usepackage{bm}
\usepackage{graphicx}
\usepackage{url}
\usepackage{savesym}
\usepackage[T1]{fontenc}
\usepackage{orcidlink}

\makeatletter
\renewcommand*\aa@pageof{, page \thepage{} of \pageref*{LastPage}}
\makeatother

\usepackage{natbib,twoopt}
\bibpunct{(}{)}{;}{a}{}{,} 

\usepackage[caption=false]{subfig}

\definecolor{green1}{RGB}{0, 128, 0}

\begin{document} 

   \title{Vertical shear instability in two-moment 
radiation-hydrodynamical simulations of irradiated protoplanetary disks}

 \titlerunning{VSI in irradiated PPDs II. Secondary instabilities and stability regions}
 \authorrunning{Melon Fuksman, Flock \& Klahr}

   \subtitle{II. Secondary instabilities and stability regions}

     \author{Julio David Melon Fuksman
          \inst{1}\orcidlink{0000-0002-1697-6433}
          \and
          Mario Flock
          \inst{1}\orcidlink{0000-0002-9298-3029}
          \and
          Hubert Klahr
          \inst{1}\orcidlink{0000-0002-8227-5467}
          }

   \institute{Max Planck Institute for Astronomy, K\"onigstuhl 17, 69117 Heidelberg, Germany\\\email{fuksman@mpia.de}}


   \date{Received 31 March 2023 /
         Accepted 30 November 2023}

\abstract
{The vertical shear instability (VSI) is a hydrodynamical instability predicted to produce turbulence in magnetically inactive regions of protoplanetary disks. The regions in which this instability can occur and the physical phenomena leading to its saturation are a current matter of research.
}
{We explore the secondary instabilities triggered by the nonlinear evolution of the VSI and their role in its saturation. We also expand on previous investigations on stability regions by considering temperature stratifications enforced by stellar irradiation and radiative cooling, and including the effects of dust-gas collisions and molecular line emission.
}
{We modeled the gas-dust mixture in a circumstellar disk around a T Tauri star by means of high-resolution axisymmetric radiation-hydrodynamical simulations including stellar irradiation with frequency-dependent opacities, considering different degrees of depletion of small dust grains.
}
{The flow pattern produced by the interplay of the axisymmetric VSI modes and the baroclinic torque forms bands of nearly uniform specific angular momentum. In the high-shear regions in between these bands, the Kelvin-Helmholtz instability (KHI) is triggered. 
A third instability mechanism, consisting of an amplification of eddies by baroclinic torques, forms meridional vortices with Mach numbers up to $\sim 0.4$. 
Our stability analysis suggests that 
protoplanetary disks can be VSI-unstable in surface layers up to tens of au for reasonably high gas emissivities.}
{The significant transfer of kinetic energy to small-scale eddies produced by the KHI and possibly even the baroclinic acceleration of eddies limit the maximum energy of the VSI modes, likely leading to the saturation of the VSI.
Depending on the gas molecular composition, the VSI can operate 
at surface layers even in regions where the midplane is stable.
This picture is consistent with current observations of disks showing thin midplane millimeter-sized dust layers while appearing vertically extended in optical and near-infrared wavelengths.
}

\keywords{protoplanetary disks  – instabilities – radiative transfer – hydrodynamics – methods: numerical}

\maketitle

\input{introduction}
\input{numericalmethod}

\input{secondaryinst}
\input{baroclinicampl}
\input{longtime}

\input{stability1}

\input{stability2}

\input{discussion}
\input{conclusions}

\begin{acknowledgements}
We thank Marcelo Barraza-Alfaro, Remo Burn, Riccardo Franceschi, Dhruv Muley, and Bhargav Vaidya for useful comments and discussions that helped improve this manuscript. The research of J.D.M.F. and H.K. is supported by the German Science Foundation (DFG) under the priority program SPP 1992: "Exoplanet Diversity" under contract KL 1469/16-1/2. We thank our collaboration partners on this project in Kiel, Sebastian Wolf and Anton Krieger, under contract WO 857/17-1/2, for providing us with the employed tabulated opacity coefficients. M.F. acknowledges funding from the European Research Council (ERC) under the European Union’s Horizon 2020 research and innovation program (grant agreement No. 757957). All numerical simulations were run on the ISAAC and VERA clusters of the MPIA and the COBRA cluster of the Max Planck Society, all of these hosted at the Max-Planck Computing and Data Facility in Garching (Germany). We thank the anonymous referee for many insightful comments that helped us improve the quality and presentation of this work.
\end{acknowledgements}

\bibliographystyle{aa}
\bibliography{refs}

\begin{appendix}
\input{trel}

\end{appendix}

\end{document}

%% file: introduction.tex
\section{Introduction}\label{S:Introduction}

The vertical shear instability \citep[VSI, ][]{Goldreich1967,UrpinBrandenburg1998} is one of several proposed hydrodynamical (HD) instabilities able to produce turbulence in weakly ionized zones of protoplanetary disks where magnetic instabilities are expected to be suppressed \citep[see, e.g.,][]{Lesur2022PPVII}. Even though numerical models predict VSI-induced turbulence to lead to negligible stellar accretion compared to current estimated rates \citep[][]{Hartmann1998,Manara2016}, this phenomenon should still produce significant vertical stirring of dust grains \citep[][]{Stoll2016,Flock2017RadHydro}, thereby increasing their collisional velocities and diffusivities and directly impacting their ability to coagulate and form planetesimals \citep{Ormel2007,Johansen2014,Klahr2020}. Furthermore, the VSI has been shown in 3D simulations to lead to the formation of small- and large-scale anticyclonic vortices \citep[][]{Richard2016,Manger2018,Pfeil2021}, which should locally concentrate dust particles, likely accelerating the formation of planetesimals \citep{Johansen2007,Gerbig2020} and potentially explaining structures observed in disks \citep[e.g.,][]{vanderMarel,deBoer2021,Marr2022}. Determining where in a disk this instability can operate is thus a necessary step toward explaining current and future observations of protoplanetary disks, as well as improving current models of planet formation.

So far, no direct observations have been made that confirm the occurrence of VSI in protoplanetary disks. As proposed by \cite{BarrazaAlfaro2021}, large-scale velocity perturbations produced by the VSI may be detectable in CO kinematics observations with the current capabilities of the Atacama Large Millimeter/submillimeter Array (ALMA). On the other hand, current ALMA observations of razor-thin millimeter-dust distributions in disks up to $\sim 100$ au \citep{Pinte2016,Doi2021,Villenave2022} seem to contradict a possible VSI activity close to the disk midplane at such distances, as the vertical stirring of grains would lead to thicker dust distributions than observed \citep[][]{Dullemond2022}. On the theoretical side, predictions of VSI-stable and unstable regions have been made \citep{Malygin2017,Pfeil2019,Fukuhara2021,Fukuhara2023} by locally applying the global stability criterion by \cite{LinYoudin2015} and using simplified prescriptions to obtain temperature distributions. One of the goals of this work is then to improve on these investigations by considering disk models with realistic temperature distributions obtained via radiative transfer, taking into account our previous analysis of local and global stability criteria \citep[][Paper I henceforth]{MelonFuksman2023}.

Besides the question of the disk stability, little is known about the mechanisms that lead to the saturation of the VSI. In \cite{LatterPapaloizou2018}, it was shown via a local Boussinesq analysis of nonlinear perturbations that, once the VSI modes reach high enough velocities, parasitic Kelvin-Helmholtz (KH) modes should be produced, limiting the maximum velocities reached by the VSI modes and eventually resulting in their saturation. Even though KH eddies are, in general, not resolved in hydrodynamical simulations, perturbations in between adjacent VSI-induced flows that could be caused by the Kelvin-Helmholtz instability (KHI) can typically be seen in both 2D and 3D global simulations \citep[e.g.,][]{Nelson2013,Manger2018,Pfeil2021}. On the other hand, an alternative saturation mechanism was recently introduced in \cite{CuiLatter2022}, where it was proposed that VSI body modes can saturate by transferring energy to small scales via resonance with pairs of inertial waves produced by nonlinear interaction (parametric instability). This instability has not yet been seen in global VSI simulations, possibly due to the expected $\gtrsim 30$ grid points per wavelength required to resolve it, estimated in that work to correspond to $\sim 300$ cells per scale height (these estimates, however, depend on the wavelength of the VSI modes).
It is then clear that very high resolutions are required to get a better understanding of the nonlinear behavior and saturation mechanism of the VSI. Such resolution requirements can be excessive in 3D hydrodynamical simulations, but they can be achieved in axisymmetric simulations at the cost of neglecting non-axisymmetric phenomena. Given that the driving mechanism of the VSI is axisymmetric, so are approximately VSI modes in 3D \citep[e.g.,][]{Flock2020,Pfeil2021,BarrazaAlfaro2021}, and thus useful information on the nonlinear evolution of the instability can be extracted from high-resolution 2D simulations. However, saturation in real disks might also be linked to the formation of non-axisymmetric structures, such as small- and large-scale anticyclonic vortices \citep[][]{Richard2016,Manger2018,Pfeil2021}.

In this work we investigated the growth and evolution of secondary instabilities triggered by the VSI in the axisymmetric radiation-hydrodynamical (Rad-HD) simulations of protoplanetary disks in Paper I and analyzed their connection to the saturation of the VSI. We also extended the local stability analysis considered in that work by including variations of the local thermal relaxation timescale in regions where either the dust-gas collisional timescale or the radiative emission by gas molecules become important. We then applied this analysis to make predictions of different stability regions in our disk models considering varying degrees of dust depletion, and we linked our resulting stability maps to current and future observations of protoplanetary disks.

This article is organized as follows. In Section \ref{S:DiskModels} we summarize the disk models and numerical methods employed in this work. In Section \ref{S:SecondaryInstabilities} we characterize the secondary instabilities triggered by the VSI in our simulations. In Section \ref{S:StabilityAnalysis} we extend the stability analysis in Paper I and produce stability maps for our disk models. In Section \ref{S:Discussion} we discuss caveats and further consequences of our results, focusing in particular on possible saturation mechanisms of the VSI and connecting our predicted stability regions with current observations of protoplanetary disks. Finally, in Section \ref{S:Conclusions} we summarize our conclusions. Additional calculations are included in the appendices.





%% file: numericalmethod.tex
\begin{table*}[t!]
\centering
\caption{Varying parameters of the simulations analyzed in this work.}
\label{tt:RadHD}
\begin{tabular}{
*{1}{p{0.11\linewidth}}
*{1}{p{0.05\linewidth}}
*{1}{p{0.11\linewidth}}
*{1}{p{0.05\linewidth}}
*{1}{p{0.07\linewidth}}
*{4}{p{0.10\linewidth}}}
\hline\hline
Label              & $f_\mathrm{dg}$ & $N_r\times N_\theta$ & $H/\Delta r$ & $t_f\,(T_{5.5})$ & $\Omega t_\mathrm{cool}(\frac{\Gamma H}{2})$ & $\Omega t_\mathrm{cool}(z_\mathrm{max})$ & $\frac{t_\mathrm{cool}}{t_\mathrm{crit}}(\frac{\Gamma H}{2})$ & $\frac{t_\mathrm{cool}}{t_\mathrm{crit}}(z_\mathrm{max})$  \\ \hline
\sftw{dg3c4\_512}   & $10^{-3}$       & $480\times 512$       & $50$   & $3800$ &  $0.01$ & $10^{-4}$ & $0.15$ & $4\times 10^{-3}$ \\
\sftw{dg4c4\_512}   & $10^{-4}$       & $480\times 512$       & $50$   & $300$ & $0.13$ & $10^{-3}$ & $1.98$ & $4\times 10^{-2}$ \\
\sftw{dg3c4\_1024}  & $10^{-3}$       & $960\times 1024$      & $100$   & $1450$ &  $0.01$ & $10^{-4}$ & $0.15$ & $4\times 10^{-3}$ \\
\sftw{dg4c4\_1024}  & $10^{-4}$       & $960\times 1024$      & $100$   & $300$ & $0.13$ & $10^{-3}$ & $1.98$ & $4\times 10^{-2}$ \\
\sftw{dg3c4\_2048}  & $10^{-3}$       & $1920\times 2048$     & $200$   & $300$ &  $0.01$ & $10^{-4}$ & $0.15$ & $4\times 10^{-3}$ \\
\sftw{dg4c4\_2048}  & $10^{-4}$       & $1920\times 2048$     & $200$   & $300$ & $0.13$ & $10^{-3}$ & $1.98$ & $4\times 10^{-2}$ \\\hline
\end{tabular}
\tablefoot{Dust-to-gas mass ratio of small grains ($f_\mathrm{dg}$), resolution ($N_r\times N_\theta$), approximate number of cells per scale height ($H/\Delta r$), and total simulation time ($t_f$) in units of the Keplerian orbital time at $5.5$ au ($T_{5.5}$). Also shown for reference are the normalized radiative cooling time $\Omega t_\mathrm{cool}$ and the ratio $\frac{t_\mathrm{cool}}{t_\mathrm{crit}}$ between the cooling time and the local critical value for instability in Equation \eqref{Eq:tcrit_loc} (Section \ref{SS:VSIcooling}), shown at the maximum height in the domain ($z/r\approx0.3$) and at the height $z=\frac{\Gamma H}{2}$ determining the vertically global stability criterion in Equation \eqref{Eq:tcrit_glob}.}
\end{table*}
 
\section{Disk models}\label{S:DiskModels}

 In this work we consider the same two disk models as in Paper I and employ the same numerical scheme therein. We solve the Rad-HD equations by means of the Rad-HD module by \cite{MelonFuksman2021} implemented in the open-source \sftw{PLUTO} code \citep[version 4.4,][]{Mignone2007}). We use spherical coordinates $(r,\theta)$ with axial symmetry, solving as well for the azimuthal $\phi$ component of all vector fields. Some quantities in this work are instead expressed in cylindrical coordinates $(R,z)$. We model the gravitational potential and heating produced by a T Tauri star of mass $M_s=0.5 M_\odot$, effective temperature $T_s=4000$ K, and radius $R_s=2.5$ $R_\odot$. The heating rate of the disk gas-dust mixture due to stellar irradiation is computed via ray-tracing using frequency-dependent absorption opacities taken from \citep{Krieger2020,Krieger2022}, assuming small dust grains dynamically well coupled to the gas with radii between $5$ and $250$ nm. The frequency-averaged values of these opacities are used for the computation of absorption and emission of reprocessed radiation.
 
 Initial hydrostatic conditions are obtained by iterating the computation of temperature and density distributions via the radiative transfer method in the code and a numerical solution of hydrostatic equations, as detailed in \cite{MelonFuksman2022}. The gas column density is in every case kept as $\Sigma(r)=600\,\mathrm{g}\,\mathrm{cm}^{-2}\,(r/\mathrm{au})^{-1}$. Once initial conditions are obtained in the domain $(r,\theta)\in[0.4,100]\,\mathrm{au}\times[\pi/2-0.5,\pi/2+0.5]$, the Rad-HD equations are solved in the smaller domain $(r,\theta)\in[4,7]\,\mathrm{au}\times[\pi/2-0.3,\pi/2+0.3]$.

All presented simulations are labeled in Table \ref{tt:RadHD} together with their corresponding parameters. For the computation of absorption opacities for both stellar irradiation and radiative transfer of reprocessed infrared photons, we assume a constant dust-to-gas mass ratio
$f_\mathrm{dg}$ of small ($<0.25$ $\mu$m) grains. We consider two $f_\mathrm{dg}$ values, representing a nominal case with $f_\mathrm{dg}=10^{-3}$ and a dust-depleted case with $f_\mathrm{dg}=10^{-4}$. Assuming a total dust-to-gas mass ratio including all grain sizes of $\rho_d^\mathrm{tot}/\rho=10^{-2}$, these values correspond respectively to $10\%$ and $1\%$ of the total dust mass being in the small-grain population, which for the grain size distribution $dn\sim a^{-3.5} da$ assumed for the opacity computation correspond to maximum grain sizes of $19$ $\mu$m and $1.8$ mm, respectively. The chosen disk region is either vertically thick or thin and has different stability regions depending on the dust content: while in the nominal case the entire domain is VSI-unstable, in the dust-depleted case the disk is stable at the middle layer located below the irradiation surfaces, defined as the regions of unity optical depth for photons emitted at the star. As detailed in Paper I, this results from the different space-dependent cooling timescales in the domain for varying dust content (see also Section \ref{S:StabilityAnalysis}).

Radiative transfer of reprocessed photons is modeled by evolving the frequency-integrated radiation energy density and flux employing the M$1$ closure by \cite{Levermore1984M1}. To increase the largest possible time step, the disparity between the radiation and HD timescales is reduced by employing an artificially small value $\hat{c}<c$ of the speed of light to advect the radiation fields. In this work we take $\hat{c}=10^{-4} c$, which, as verified in Paper I via analytical and numerical testing, does not introduce unphysical phenomena in our simulations. Together with the employment of a reduced radial domain compared to the disk size, this allows us to achieve the high resolutions needed to investigate secondary instabilities. We employ a maximum resolution of $N_r\times N_\theta=1920\times 2048$, which corresponds to approximately $200$ cells per pressure scale height $H$ (see Table \ref{tt:RadHD}) with an aspect ratio of $1:1$. We limit numerical diffusion by employing third-order WENO spatial reconstruction \citep{Yamaleev2009,Mignone2014reconstruction} and HLLC-type Riemann solvers for both radiation \citep{MelonFuksman2019} and matter fields \citep{Toro}. Dirichlet boundary conditions are applied by fixing all fields to the hydrostatic initial conditions at the ghost zones. Unless otherwise stated, all results presented in this work correspond to our highest-resolution runs.

%% file: secondaryinst.tex
\section{Secondary instabilities}\label{S:SecondaryInstabilities}


\subsection{Overview}\label{SS:SecondaryInstOverview}

\begin{figure*}[t!]
\centering
\includegraphics[width=\linewidth]{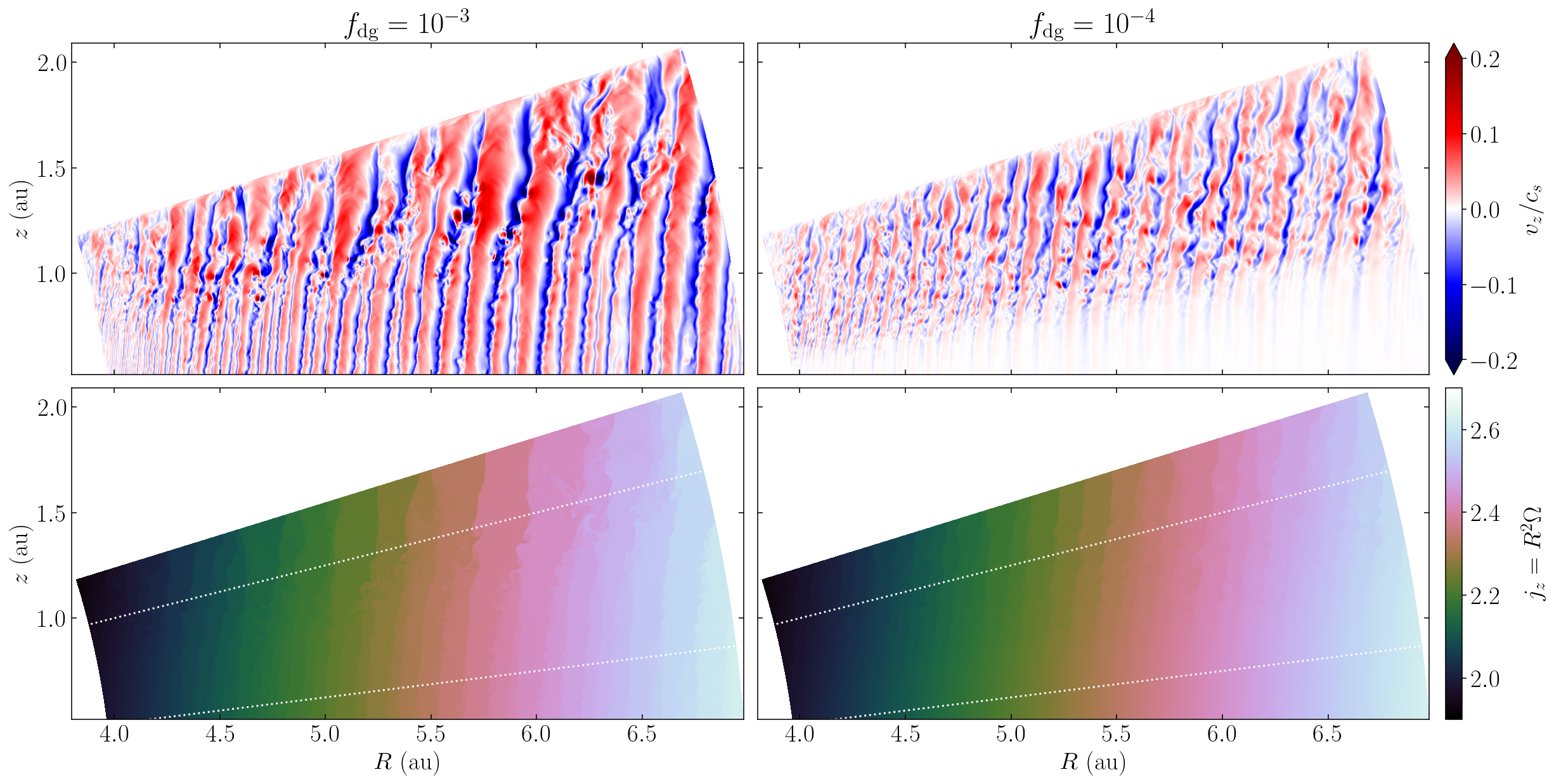}
\caption{Vertical velocity normalized by the local sound speed (top) and specific angular momentum $j_z=R^2\Omega$ (bottom) in the upper region of our highest-resolution simulations at $t=234$ $T_{5.5}$. White dotted lines indicate the $z/r=0.125$ and $0.25$ slices displayed in Fig. \ref{fig:jz_zr}.}
\label{fig:veljz}
\end{figure*}

\begin{figure}[t!]
\centering
\includegraphics[width=\linewidth]{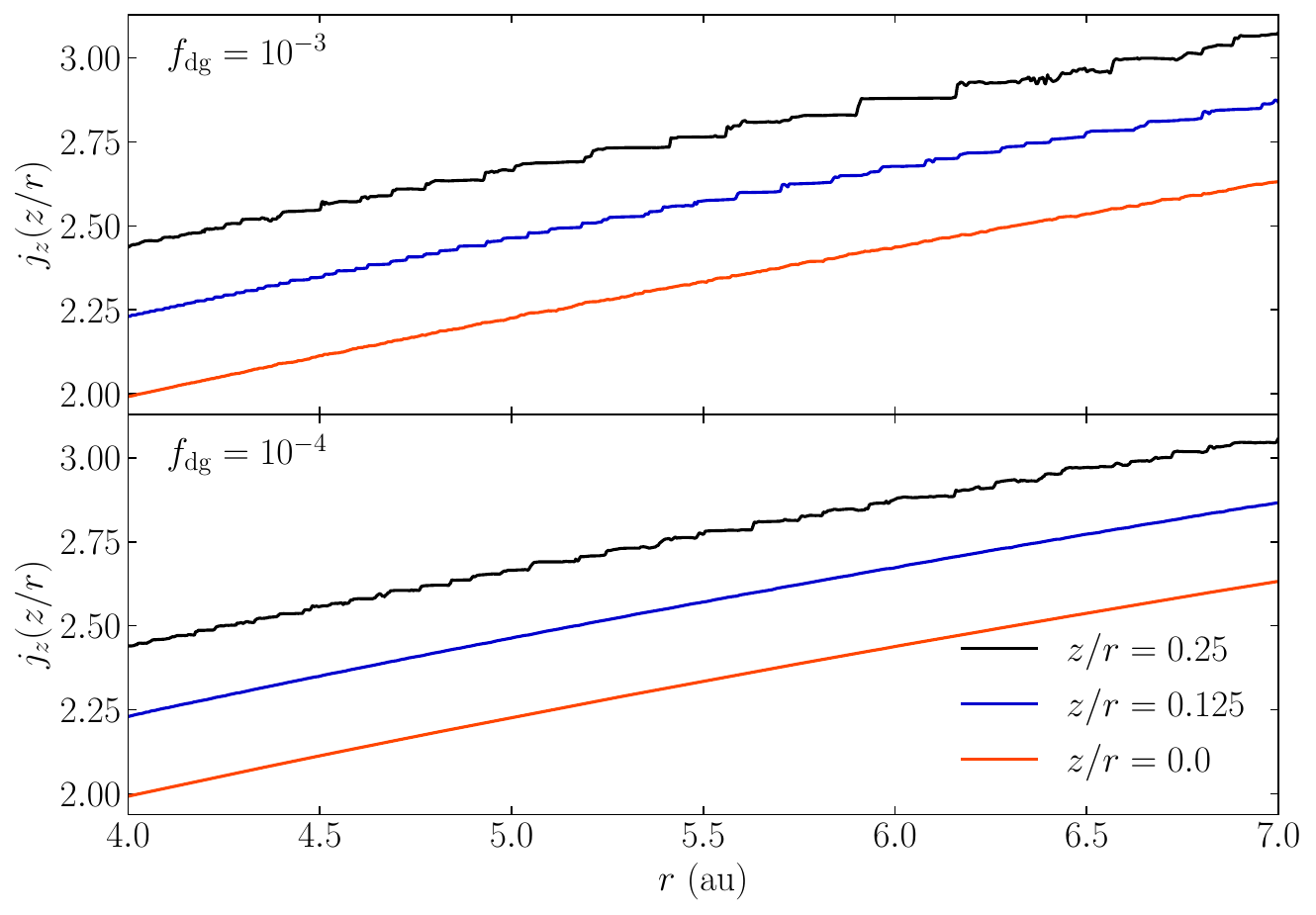}
\caption{Specific angular momentum at fixed $z/r$ values in Fig. \ref{fig:veljz}. The curves have have been vertically shifted proportionally to $z/r$ for better visualization.}
\label{fig:jz_zr}
\end{figure}

We begin by summarizing the main ideas presented in this section, which comprise our main results regarding secondary instabilities resulting from the evolution of the axisymmetric VSI. In its linear stage, the VSI transports gas in nearly vertical directions. Since the unstable directions of motion cross surfaces of constant specific angular momentum $j_z=R^2\Omega$, this leads to angular momentum redistribution. As shown in Paper I, the nonlinear development of the VSI results in the formation of nearly uniform-$j_z$ bands, in which the initial vertical shear is significantly reduced. This can be seen in Figs. \ref{fig:veljz} and \ref{fig:jz_zr}, showing velocity and $j_z$ distributions after VSI saturation.

As shown in these figures and schematized in Fig. \ref{fig:KHdiagram}, the gas inside the constant-$j_z$ bands rotates clockwise above the midplane (in a reference frame in which the orbital velocity enters the meridional plane) and counterclockwise below the midplane, with minimum speed ad the center of the bands and maximum speed at the edges. Due to the abrupt reversal of the vertical velocity at the interfaces between adjacent bands, significant meridional shear is created at such regions, eventually triggering the KHI. While this instability typically involves a balance of the destabilizing effect of shear and the stabilizing effect of buoyancy (e.g., in atmospheric physics), the latter is suppressed in our considered system by the fast cooling required for the VSI to operate. Instead, stabilization results from the radial angular momentum stratification, and a KH-stability criterion can be written in terms of a Richardson number in which the Brunt-Vais\"al\"a frequency is replaced by the epicyclic frequency (Section \ref{SS:KH}). The VSI growth phase ends as soon as the KH eddies comprise a significant portion of the kinetic energy, supporting the hypothesis that the KHI is the main mechanism behind VSI saturation in disks (Section \ref{SS:SpectralAnalysis}).

Various aspects of both the large-scale structure of the vertical flows in the nonlinear stage of the VSI and the amplification of small-scale eddies in our simulations can be explained in terms of the generation of vorticity in rotating baroclinic flows. Locally, the time evolution of the azimuthal vorticity $\omega_\phi=(\nabla\times\mathbf{v})_\phi$ depends on a combination of three terms, denoted $\tau_a$, $\tau_b$, and $\tau_c$ (Section \ref{SS:Vortices}). The terms $\tau_b$ and $\tau_c$ account for baroclinic and centrifugal torques, respectively, while $\tau_a$ accounts for vorticity advection. Once the VSI starts growing, $\tau_b$ starts overcoming $\tau_c\propto\partial_z j_z^2$ as constant-$j_z$ regions start forming, contributing to the clockwise rotation of the large-scale flows at $z>0$ and counterclockwise rotation at $z<0$. Although the term $\tau_a$ is generally dominant over the other two, the acceleration or deceleration of vortices advected with the flow only depends on the balance of $\tau_b$ and $\tau_c$. Thus, in constant-$j_z$ regions, $\tau_b$ is the dominant term contributing to vortex acceleration, which results in an acceleration of vortices with $\sgn{\omega_\phi}=\sgn{\tau_b}$ and, vice versa, a deceleration of vortices with $\sgn{\omega_\phi}=-\sgn{\tau_b}$. Even though the KHI mainly produces vortices rotating in the direction disfavored by the baroclinic torque, its nonlinear evolution also results, via $\tau_a$, in the formation of counter-rotating vortices in regions with $\sgn{\omega_\phi}=\sgn{\tau_b}$. These vortices can then be accelerated by $\tau_b$ in the constant-$j_z$ bands. Since $\sgn{\tau_b}=\sgn{z}$, this leads to the formation of small-scale vortices rotating clockwise at $z>0$ and counterclockwise at $z<0$, which are accelerated up to significant fractions of the local sound speed (Sections \ref{SS:Vortices} and \ref{SS:MeridionalVorticesAmpl}). Further acceleration of these vortices is likely produced via advection from regions adjacent to the shear layers with $\sgn{\omega_\phi}=\sgn{\tau_b}$. The size of these vortices is limited by the width of the constant-$j_z$ regions. In axisymmetric simulations, in which such regions can grow in size unimpeded by non-axisymmetric instabilities, the baroclinically amplified vortices can eventually grow up to large sizes ($\sim 1$ au) dominating the gas dynamics, as detailed in Section \ref{A:LongTime}.

\subsection{Kelvin-Helmholtz instability}\label{SS:KH}

\begin{figure}
\centering
\includegraphics[width=\linewidth]{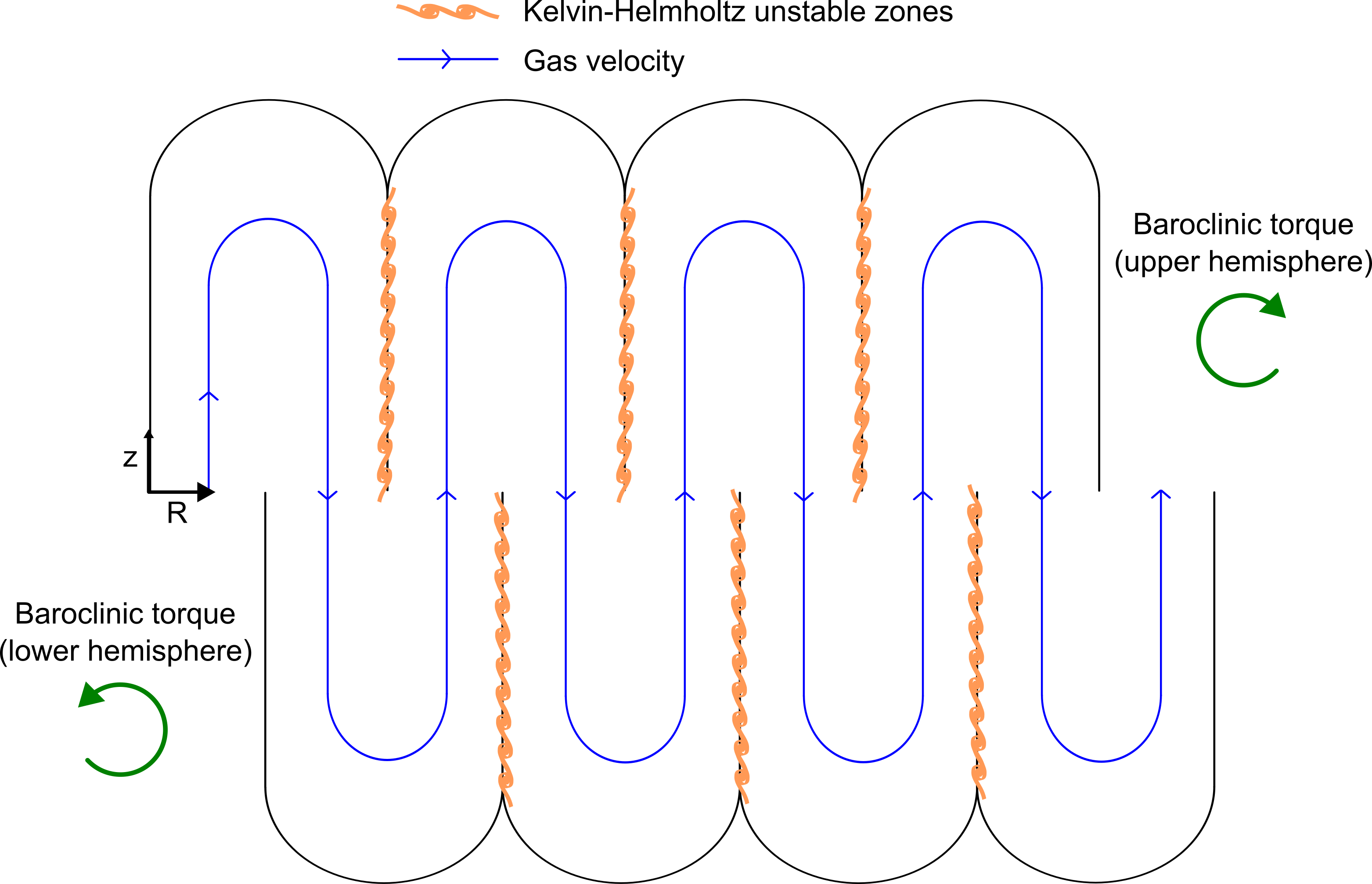}
\caption{Schematic view of the KH-unstable zones in a VSI-unstable axisymmetric disk. Gas rotates in the meridional plane inside of approximately constant-$j_z$ zones in the sense of rotation favored by the baroclinic torque, producing high-shear regions that become KH-unstable. The transfer of kinetic energy to KH eddies acts as an effective friction between the vertical flows and contributes to the saturation of the VSI.}
\label{fig:KHdiagram}
\end{figure}
\begin{figure}[t!]
\centering
\includegraphics[width=\linewidth]{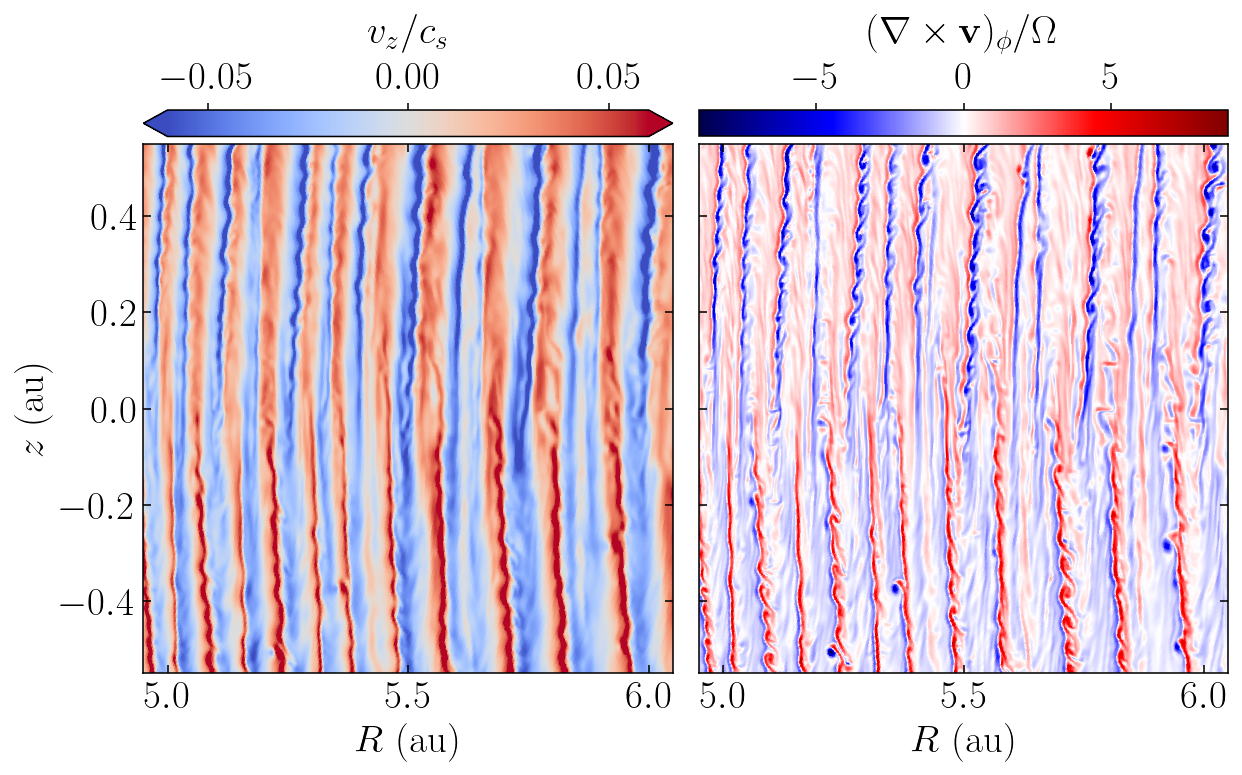}
\caption{Velocity and vorticity distributions in a region close to the midplane in run \sftw{dg3c4\_2048} after $300$ orbits. Red and blue regions in the vorticity map correspond to clockwise and counterclockwise rotation, respectively. KH eddies on the lower hemisphere rotate clockwise, while their nonlinear evolution produces counterclockwise-rotating eddies that are amplified by the baroclinic torque, visible in that region as dark blue spots.}
\label{fig:KHsim}
\end{figure}
\begin{figure}[t!]
\centering
\includegraphics[width=\linewidth]{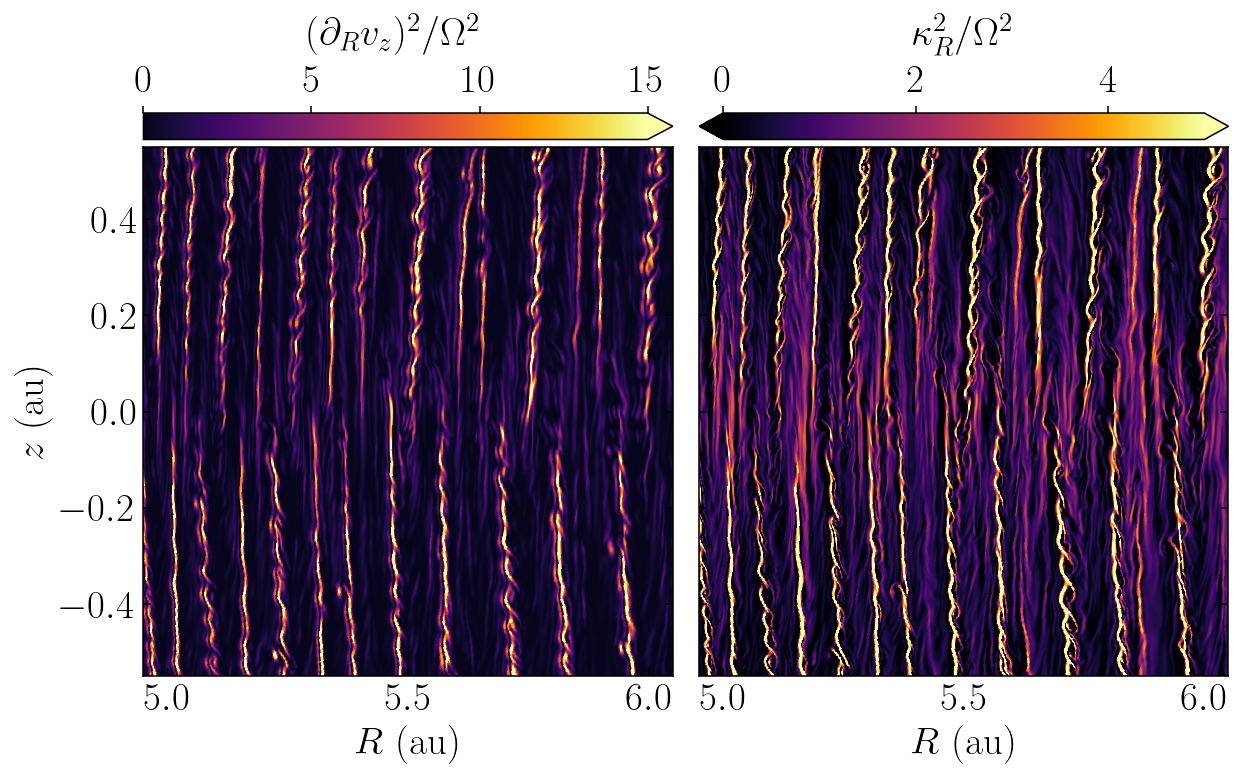}
\caption{Same as Fig.  \ref{fig:KHsim}, showing instead $(\partial_R v_z)^2$ and the squared epicyclic frequency $\kappa_R^2$, both normalized by the squared rotation angular velocity.}
\label{fig:KHsimRichardson}
\end{figure}

We begin by analyzing the triggering of the KHI in between uniform-$j_z$ regions.
As shown in Figs. \ref{fig:veljz}, \ref{fig:KHdiagram}, and \ref{fig:KHsim}, the rotation pattern inside of the approximately uniform-$j_z$ bands causes downward-moving gas parcels at $z>0$ to experience significant shear on their larger-$r$ edges, while below the midplane shear is maximum on the smaller-$r$ edges. For $f_\mathrm{dg}=10^{-3}$, in which case the VSI flows cross the midplane, this produces an intercalated pattern of KH-unstable regions, each of them occupying half of the vertical domain (see also Fig. \ref{fig:KHsimRichardson}). The same occurs for $f_\mathrm{dg}=10^{-4}$ in the upper VSI-unstable layers.
The resulting eddies are clearly highlighted by $\omega_\phi$, which also shows their sense of rotation. We show this quantity in Fig. \ref{fig:KHsim} next to the normalized vertical velocity in a region close to the midplane for run \sftw{dg3c4\_2048}. 

To operate, the KHI requires that the kinetic energy available to displace a fluid element across a shear layer overcomes the counteracting work of restoring forces. This balance is generally quantified in terms of the Richardson number \citep[e.g.,][]{Chandrasekhar1961}, defined as the ratio between these energies. While typically (e.g. in shear flows perpendicular to gravity in the atmosphere or the ocean) buoyancy provides the restoring force, this effect is in our case suppressed by fast cooling. Conversely, the increase of $j_z$ with radius contributes to the stabilization of radial displacements. In light of these considerations, we define the radial Richardson number as 
\begin{equation}\label{Eq:Richardson}
\mathrm{Ri}=\frac{\kappa_R^2}{(\partial_R v_z)^2}\,,
\end{equation}
where $\kappa_R^2=\frac{1}{R^3}\partial_R j_z^2$ is the local epicyclic frequency. To obtain this expression, we have simply replaced the squared restoring frequency in the numerator, typically given by the buoyancy frequency \citep[e.g.,][]{Chandrasekhar1961}, with the squared frequency $\omega^2$ of radially oscillating parcels in stratified disks, given in \citep[][]{Goldreich1967,Urpin2003,Klahr2023a}. While in the adiabatic limit $\omega^2 = N_R^2+\kappa_R^2$, where $N_R$ is the radial Brunt-Vais\"al\"a frequency, $\omega^2$ tends to $\kappa_R^2$ for fast thermal relaxation compared to epicyclic motion ($t_\mathrm{cool}\ll \Omega^{-1}$, as shown in Section \ref{SS:VSIcooling}), leading to Eq. \eqref{Eq:Richardson}.
In the context of rotating stars, an equivalent expression is obtained in \cite{Maeder2013} in the limits of instant cooling and uniform molecular weight.

\begin{figure*}[t!]
\centering
\includegraphics[width=\linewidth]{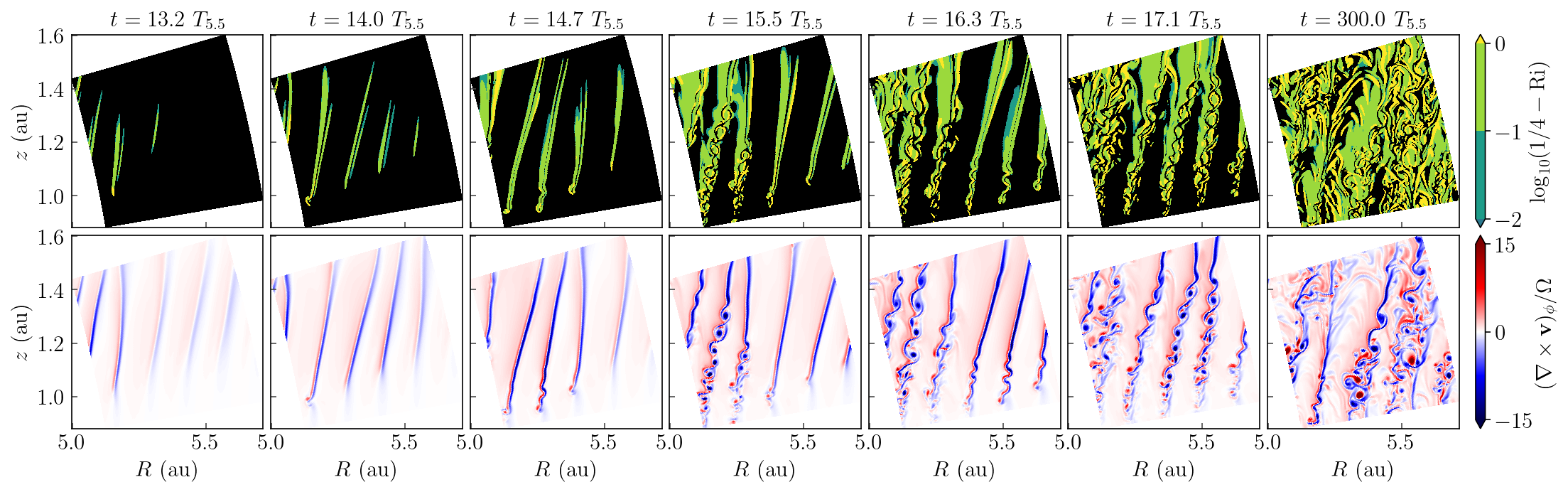}
\caption{Onset of the KHI in run \sftw{dg3c4\_2048} in a region at the disk upper layers, where the instability first occurs. Top: snapshots at different times (in units of the orbital period at $5.5$ au) showing the deviation of the radial Richardson number (Eq. \eqref{Eq:Richardson}) from its critical value of $1/4$ required to trigger the KHI. Black regions correspond to $\mathrm{Ri}>1/4$. Bottom: same snapshot showing the azimuthal vorticity, highlighting the formation of eddies. The rightmost panels show the same quantities $>200$ orbits after VSI saturation.}
\label{fig:KHsimRichardson2}
\end{figure*}

A necessary condition for KH-stability is $\mathrm{Ri}<1/4$ \citep[][]{Chandrasekhar1961,Maeder2013}.
As soon as VSI modes are formed, this condition is only violated in the high-shear regions in between vertical flows (shown in Fig. \ref{fig:KHsimRichardson} in the nonlinear stage). At those locations, KH eddies are visible within a few orbits after this condition is met, as shown in Fig. \ref{fig:KHsimRichardson2}. An alternative instability criterion based on a linear analysis of perturbations to the VSI modes was derived in \cite{LatterPapaloizou2018}, requiring that $k V / \Omega \gtrsim 1.38$ to trigger the KHI, where $k$ is the wavenumber of the VSI modes and and $V$ their maximum velocity amplitude. This criterion is also consistent with our results, as we only see signs of KHI growth in regions where $k V / \Omega$ is at least of order unity (typically at least $\sim 2$), which we have verified using the wavelength distributions estimated in Paper I.

As the VSI starts forming bands of approximately uniform $j_z$, where, consequently, $\kappa_R \ll \Omega$ (Fig. \ref{fig:KHsimRichardson}), the meridional shear in the nonlinear state causes the condition $\mathrm{Ri}<1/4$ to be satisfied in the entire regions where the VSI operates (top right panel in Fig. \ref{fig:KHsimRichardson2}). Despite this, we do not observe a similar spontaneous formation of eddies inside of such bands as at their interfaces. This can possibly be explained by a lower KHI growth rate in those regions, which we expect to be proportional to the shear $\partial_R v_z$. For instance, the growth rate for a uniform fluid with a discontinuous velocity transition $\Delta v$ is $k|\Delta v|/2$ \citep[e.g.,][]{Chandrasekhar1961}, where $k$ is the component of the wavenumber parallel to the discontinuity interface. 

The VSI not only increases the shear in the meridional plane. As evidenced by the $\kappa_R$ distribution in Fig. \ref{fig:KHsimRichardson}, jumps in the specific angular momentum in between vertical flows produce regions of large azimuthal shear ($\partial_R v_\phi$). If the same happens in 3D, as it appears to be the case in the simulations by \cite{Richard2016}, \cite{Manger2018}, and \cite{Flock2020} (see Paper I), we can expect the KHI to produce azimuthal eddies in the same regions. Since $j_z$ increases with radius, such eddies should have prograde rotation with respect to the disk. Therefore, long-lived anticyclonic vortices such as those in \cite{Manger2018} and \cite{Pfeil2021}, if formed via KHI, can only result from secondary eddies produced by the nonlinear evolution of this instability in adjacent regions with a negative vorticity perturbation, similarly to the formation of counter-rotating eddies in the meridional plane (Section \ref{SS:Vortices}). 


\subsection{Energy spectra}\label{SS:SpectralAnalysis}

\begin{figure}[t!]
\centering
\includegraphics[width=\linewidth]{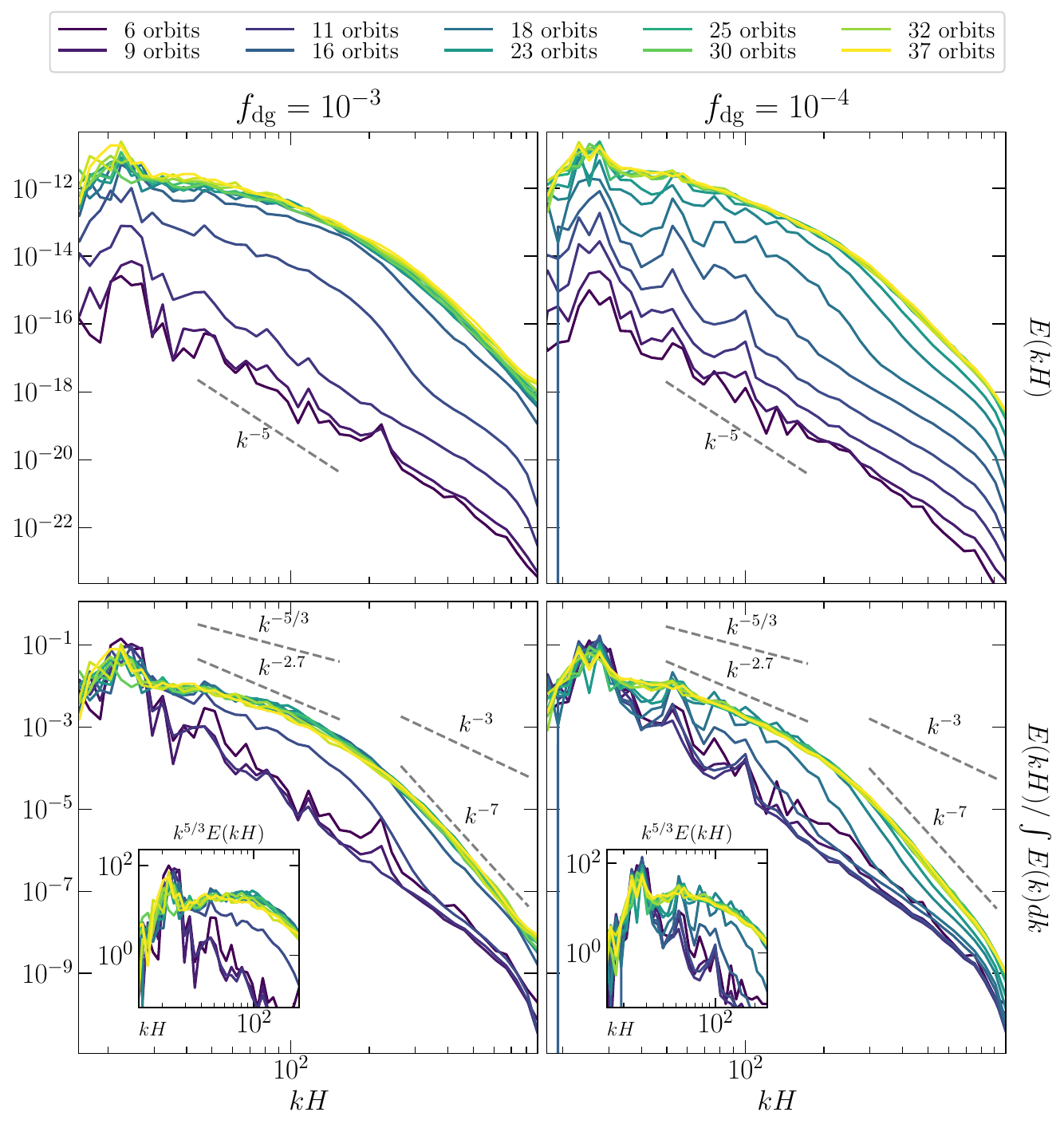}
\caption{Energy spectra. Top: Values in code units at different times computed as described in Appendix \ref{A:Fourier}. Bottom: Normalized values computed at the same times. Some relevant slopes mentioned in Sections \ref{SS:SpectralAnalysis} and \ref{SS:EnergyTransportScales} are shown for comparison.}
\label{fig:spectra}
\end{figure}

Further information linking the VSI saturation process with the onset of the KHI can be obtained by inspecting the evolution of the kinetic energy spectrum in our highest-resolution runs. As detailed in Appendix \ref{A:Fourier}, we achieve this by operating on the Fourier-transformed meridional momentum and velocity distributions in an upper disk region exhibiting VSI growth for both $f_\mathrm{dg}$ values. Energy spectra between $6$ and $37$ orbital times are shown in Fig. \ref{fig:spectra} as a function of $k H$, where $k=||(k_1,k_2)||$ is the wavenumber norm and $H$ is the average local pressure scale height in the selected region. This value is larger than at the midplane, and therefore so is the resolution relative to $H$, with $H/\Delta r\sim 263-288$ for $f_\mathrm{dg}=10^{-3}$ and $296-324$ for $f_\mathrm{dg}=10^{-4}$. The curves have been averaged every $3$ snapshots ($\sim 2.3$ orbits) to reduce noise. To highlight the energy redistribution over time, we also show in that figure the same spectra normalized by their integrated values.

As the VSI modes grow in amplitude, energy increases over time for all $k$. At the linear stage, the VSI primary modes produce an energy peak around $kH\sim 20-30$, which, after sharply decreasing at $kH\sim 30$, decays approximately as $k^{-5}$ up to $kH\gtrsim 100$. During that phase, energy at the largest wavenumbers exists due to the functional form of the developed VSI flows. In particular, as the VSI grows, the resulting flows develop sharp variations in the vicinity of the interfaces between the forming constant-$j_z$ bands.

The beginning of the saturation phase, occurring at $\sim 15-25$ orbits depending on $f_\mathrm{dg}$, coincides with a deviation from the initial spectrum's shape. At that time, a `hump' is formed at $kH\sim 50-200$ due to the formation of KH eddies in that size range. The spectra converge to a new shape at saturation, with a short plateau following the main VSI peak up to $kH\sim 50$. This plateau is followed by a shallower decrease with $k$ than in the linear growth stage, with $E(k)\sim k^{-1.7\pm 0.2}$ at $kH\sim50-100$ for $f_\mathrm{dg}=10^{-3}$ and $E(k)\sim k^{-2.7\pm 0.1}$ at $kH\sim50-200$ for $f_\mathrm{dg}=10^{-4}$. For higher wavenumbers, the spectra decay as $k^{-\varsigma}$, with $\varsigma\gtrsim 7$.

The described transition in the shape of the spectra over time  has the consequence that the proportion of the total kinetic energy contained in large scales is smaller in the saturated state than in the linear growth phase. More precisely, taking $kH=35$ as a limit between small and large scales (this value is in between the main peak produced by the VSI modes and the region dominated by KH eddies), the energy in large scales decreases between $\sim 95\%$ at the  growth stage and $50\%-60\%$ after saturation. This shows that saturation occurs as soon as KH eddies become energetically important, suggesting that some kinetic energy injected by the VSI driving mechanism is diverted into creating KH modes, eventually halting the growth of the VSI.

The obtained steepening of the spectra at $kH>100-200$ can have several possible explanations. First of all, the width of the shear layers triggering the KHI imposes a wavenumber range in which KH linear modes can grow \citep{Chandrasekhar1961}. Thus, we should expect the energy spectrum to decrease faster above the maximum wavenumber allowing for KHI growth, with a slope determined by a combination of the energy spectrum of the high-$k$ end of the VSI-induced structures and possibly a direct enstrophy cascade (a discussion on the energy transport over scales is given in Section \ref{SS:EnergyTransportScales}). The precise limits of this range typically depend on the fluid's velocity, the density stratification, and the Richardson number, but for this analysis it is enough to consider that its upper limit is some number on the order of $1/d$, where $d$ is the typical width of the shear layers \citep{Chandrasekhar1961}. In the region considered for the computation of the energy spectra, we typically have $d\sim0.02-0.03 H$, corresponding to wavenumbers $kH\sim200-300$, coinciding with the region where we see the slope transition. In this regard, we must mention that these layers are typically resolved by $\sim 10$ cells, and thus it is possible that numerical diffusion is enlarging their typical width with respect to its zero-diffusion value, artificially shifting the slope transition to lower wavenumbers.

On the other hand, numerical diffusion can also contribute to the steepening of the energy spectrum at high $k$. This occurs in the 2D homogeneous turbulence simulation made by \cite{Seligman2017} with PLUTO, also using third-order Runge-Kutta scheme and WENO3 reconstruction. In that test, numerical diffusion leads to a steeper spectrum than the predicted $E(k)\sim k^{-3}$ (for a direct enstrophy cascade) up to scales $10$ times larger than the grid's resolution. However, unlike in our simulations, no energy is continuously injected into the system, and thus the spectrum does not reach a steady state over time. Moreover, we note that our spectra steepen exhibit further steepening at $kH\sim 600-800$ (about $3$ cells per wavelength), which could indicate that the dissipation range is instead closer to this region. Yet, the relevance of numerical diffusion on the high-end of our VSI spectra can only be assessed via a proper comparison with 2D forced turbulence tests injecting energy at $k\lesssim 2\pi/d$, as we plan to do in a future study. 

\begin{figure*}[t!]
\centering
\includegraphics[width=\linewidth]{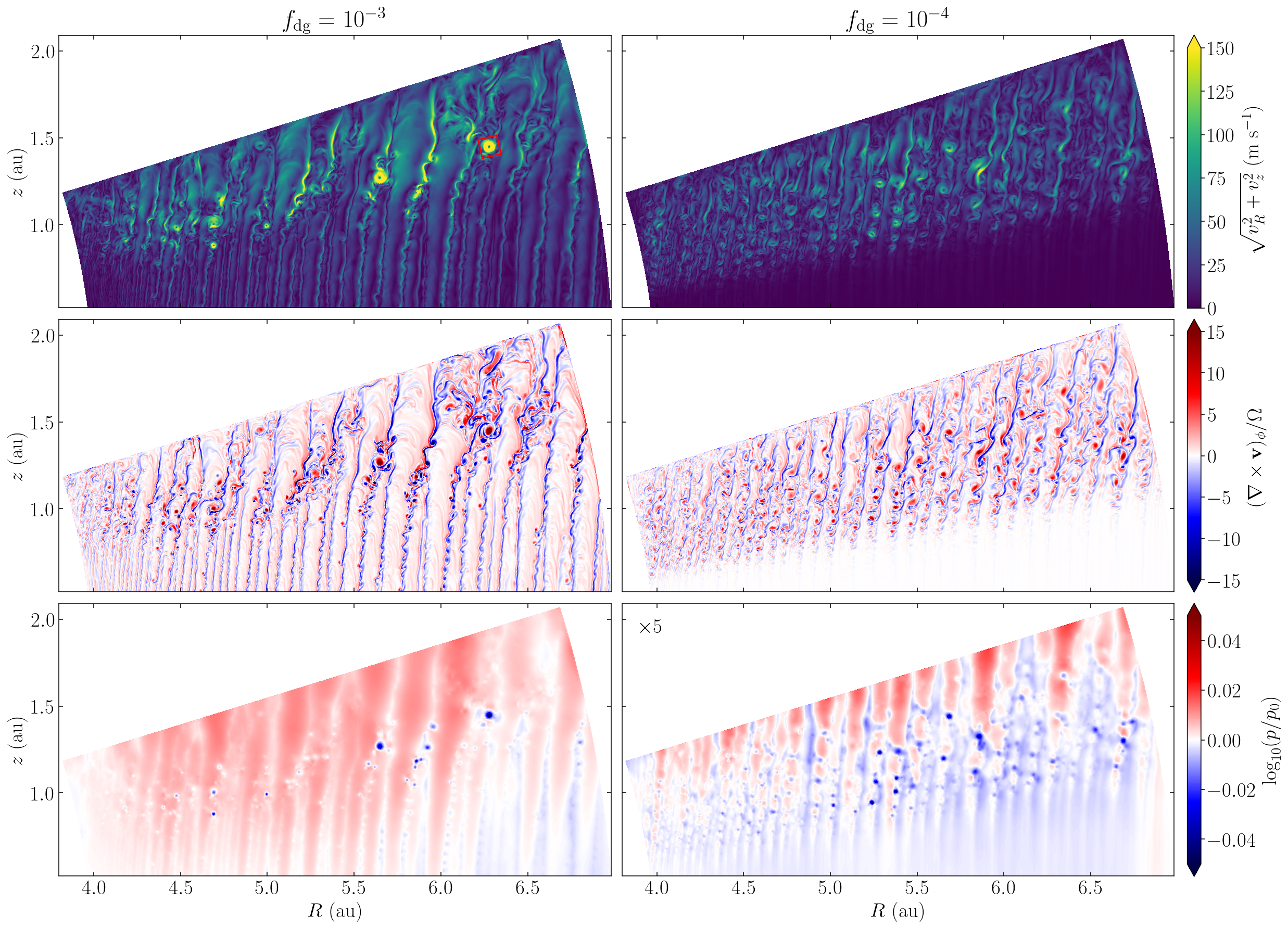}
\caption{Meridional velocity norm (top), vorticity (middle), and pressure perturbation (bottom) distributions in the upper region of our highest-resolution simulations at $t=234$ $T_{5.5}$. The $\log(p/p_0)$ values are multiplied by $5$ for $f_\mathrm{dg}=10^{-4}$ for better visualization.}
\label{fig:vort}
\end{figure*}

%% file: baroclinicampl.tex
\subsection{Vorticity generation}\label{SS:Vortices}

 In our highest-resolution simulations ($N_\theta=1024$ and $2048$), we observe a formation of small meridional vortices in the VSI-active  regions that survive up to $1-10$ orbital timescales and get accelerated up to hundreds of m s$^{-1}$, with Mach numbers up to $0.4$ for $f_\mathrm{dg}=10^{-3}$ and $0.2$ for $f_\mathrm{dg}=10^{-4}$ (see Figs. \ref{fig:vort} and \ref{fig:vortex}). These vortices, which are typically responsible for the maximum velocities reached in the domain (Fig. 5 in Paper I), rotate clockwise at $z>0$ and counterclockwise at $z<0$. Similar meridional vortices also form in the 2D isothermal and $\beta$-cooling simulations by \cite{Klahr2023b}. As argued by these authors, the observed acceleration is not produced by the convective overstability \citep[COS,][]{Klahr2014,Lyra2014}, given that the cooling timescale is much smaller than $\Omega$ in the VSI-active regions where the accelerated vortices form (Section \ref{SS:VSIcooling}). Further evidence for this is the fact that these vortices appear even in isothermal simulations, that is, in absence of buoyancy, as we also verified in isothermal versions of our setup at the same resolution (not shown here). On the other hand, the driving mechanism cannot be the VSI, since the same phenomenon occurs if the gravitational potential is modified in such a way that the disk is still baroclinic but has no vertical shear \citep{Klahr2023b}. On the contrary, the vortices thrive in bands of approximately uniform specific angular momentum, where they are instead accelerated by the meridional baroclinic torque. This is a novel instability mechanism in the context of protoplanetary disks, analogous to the sea breeze phenomenon in atmospheric physics \citep[see, e.g.,][]{Holton2013}.
 
 We can understand how the driving mechanism works by examining the time evolution of the integrated vorticity inside of closed streamlines in the $(R,z)$-plane advected with the fluid. 
 We begin by considering a closed streamline $C$ in the meridional plane defined in such a way that it is at all locations tangent to the $(R,z)$-projected velocity, $\mathbf{v}_p=v_R \bm{\hat{R}}+v_z \bm{\hat{z}}$ ($C$ is not strictly a streamline in the sense that its tangent vector is not $\mathbf{v}$ but $\mathbf{v}_p$). In Appendix \ref{A:VorticityGeneration}, it is shown that the circulation around $C$, namely, the clockwise line integral of $\mathbf{v}$ along $C$, follows the evolution equation
\begin{equation}\label{Eq:Circulation}
\begin{split}
    \frac{d}{dt} \oint_C \mathbf{v} \cdot d\bm{\ell} &= 
    \int_{S} \left(
    \tau_b+\tau_c
    \right) \, dS\,,
\end{split}
\end{equation}
where $S$ is the surface enclosed by $C$ and the time derivative takes into account the evolution of $C$ as the fluid evolves.
Equivalently, using Stokes' theorem, the circulation can be replaced as a surface integral as
\begin{equation}\label{Eq:VorticityIntegral}
    \frac{d}{dt} \int_{S} \omega_\phi  \, dS = 
    \int_{S} \left(
    \tau_b+\tau_c
    \right) \, dS\,,
\end{equation}
where $\omega_\phi=(\mathbf{\nabla}\times\mathbf{v})_\phi$. In these expressions, we have defined the baroclinic and centrifugal terms, $\tau_b$ and $\tau_c$ respectively, as
\begin{equation}
    \begin{split}
        \tau_b&=(\nabla \rho\times\nabla p)_\phi/\rho^2 \\
        \tau_c&=\kappa_z^2 \,,
    \end{split}
\end{equation}
 where  $\kappa^2_z=\frac{1}{R^3}\partial_z j_z^2=R \partial_z \Omega^2$ is the analog to $\kappa^2_R$ quantifying the vertical squared specific angular momentum gradient. The first and second terms in the right-hand side of Equations \eqref{Eq:Circulation} and \eqref{Eq:VorticityIntegral} quantify respectively the work per unit density along close curves of the centrifugal force and the expansion work. An illustrative example, taken from \cite{KunduCohen}, explains why the second term leads to the generation of vorticity: a mass element in a baroclinic fluid experiences a nonzero net torque due to the misalignment of the center of gravity with respect to the line that passes through the center of buoyancy in the direction of the net pressure force. In a similar way, a nonzero $\kappa_z^2$ produces a nonzero net torque on meridional vortices due to the mismatch of the centrifugal acceleration $R\Omega^2$ when the gas moves outward and inward in $R$.
 
\begin{figure}[t!]
\centering
\includegraphics[width=\linewidth]{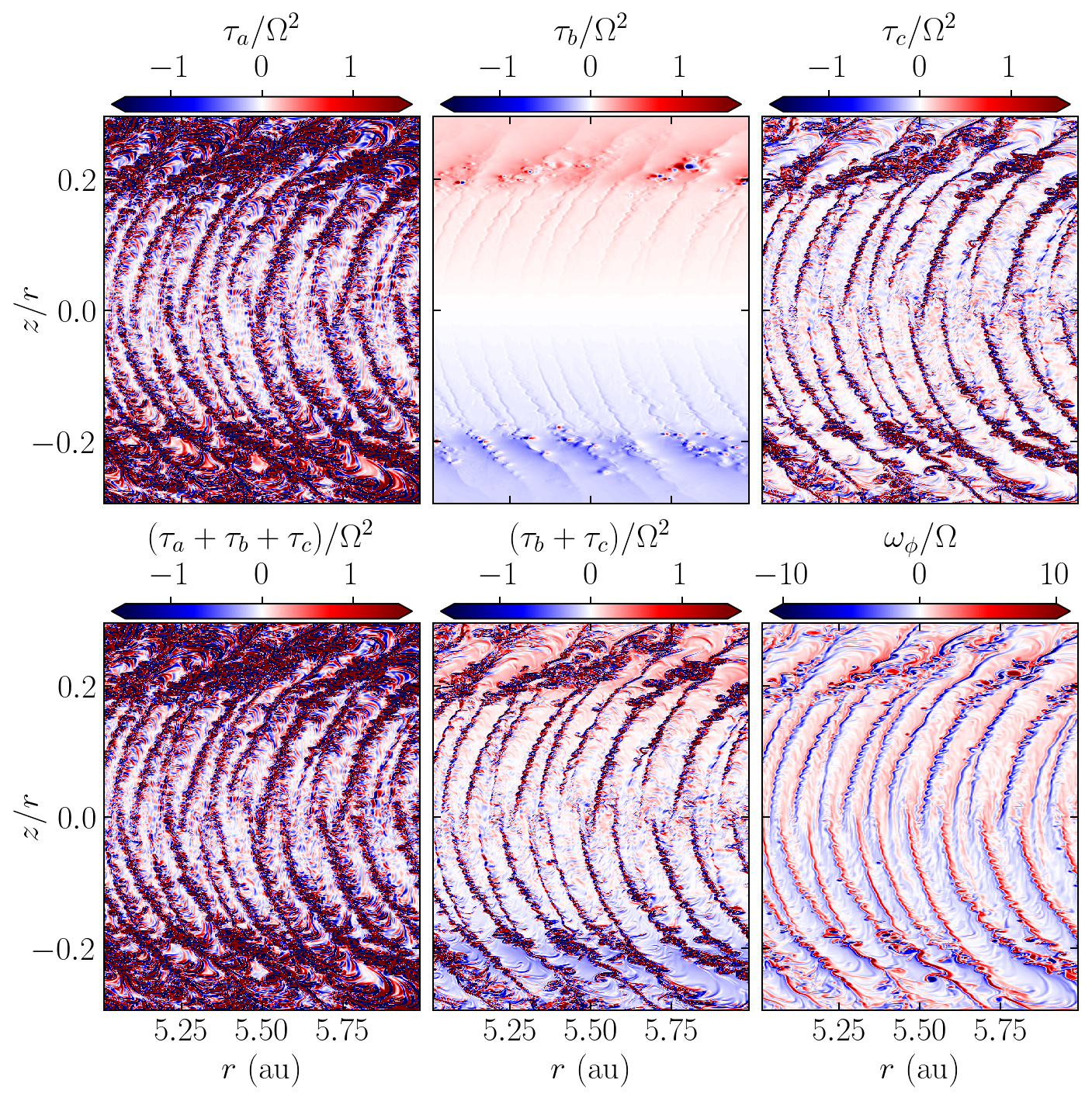}
\caption{Advection, baroclinic, and centrifugal terms ($\tau_a$, $\tau_b$, and $\tau_c$, respectively) determining the vorticity evolution (Equations \eqref{Eq:VorticityIntegral} and \eqref{Eq:Vorticity}), computed at the same snapshot as Fig. \ref{fig:vort}.}
\label{fig:vort_generation}
\end{figure}

A similar expression quantifying the vorticity evolution at fixed locations can be obtained by taking the curl of the velocity evolution equation (see Appendix \ref{A:VorticityGeneration}), namely
\begin{equation}\label{Eq:Vorticity}
    \partial_t \omega_\phi + \nabla \cdot \left(\omega_\phi\mathbf{v}_p\right)
    = \tau_b + \tau_c\,.
\end{equation}
In other words, $\omega_\phi$ can vary either by transport throughout the domain, determined by the advection term
\begin{equation}
\tau_a= - \nabla \cdot \left(\omega_\phi\mathbf{v}_p\right)\,,
\end{equation}
or by local generation, determined by the torque terms $\tau_b$ and $\tau_c$. If $\tau_b=\tau_c=0$, Eq. \eqref{Eq:Vorticity} becomes a conservation equation for the total integrated $\omega_\phi$.

In the initial hydrostatic state of our simulations, we have $\tau_a=0$ and $\tau_c+\tau_b=0$, and thus no vorticity is initially generated (the latter expression is the thermal wind equation relating the vertical shear to the disk's baroclinicity; see Equation 5 in Paper I). As shown in Fig. \ref{fig:vort_generation}, this situation changes as the VSI develops, until eventually $|\tau_a|\gg |\tau_b|, |\tau_c|$ in most of the domain. 
However, $\tau_a$ can only transport existing vorticity without creating nor destroying it. Thus, as long as vortices are advected but not destroyed via nonlinear interaction with the flow, their acceleration or deceleration depends solely on the balance of $\tau_b$ and $\tau_c$ in their enclosed surfaces (Equation \eqref{Eq:Circulation}). As soon as uniform-$j_z$ bands start to form, the term $\tau_c\propto \partial_z j_z^2$ significantly decreases in such regions until it becomes negligible compared to $\tau_b$. Therefore, the acceleration of closed streamlines transported into those regions is entirely determined by the baroclinic torque.
This means that vortices in these bands with $\sgn{\omega_\phi}=\sgn{\tau_b}$ are accelerated and, vice versa, vortices with $\sgn{\omega_\phi}=-\sgn{\tau_b}$ are decelerated. Since $\sgn{\tau_b}=\sgn{z}$, the amplified vortices in our simulations rotate clockwise for $z>0$ and counterclockwise for $z<0$.

The KHI is first triggered in shear layers with $\omega_\phi$ of the opposite sign as $\tau_b$ (see, e.g., Figs. \ref{fig:KHsim} and \ref{fig:KHsimRichardson2}) due to the spatial distribution of the VSI flows (Fig. \ref{fig:KHdiagram}). This can be easily seen by taking $v_z\gg v_R$, leading to $\omega_\phi\approx -\partial_R v_z$, and resulting in $\sgn{\omega_\phi}=-\sgn{z}$ at the shear layers. Therefore, the initially produced KH eddies can only be decelerated by $\tau_b$ if they are transported into the constant-$j_z$ zones. However, the KHI also produces via advection vortices of opposite $\omega_\phi$ as the initially produced KH eddies, some of which can be seen in Figs. \ref{fig:KHsim}, \ref{fig:KHsimRichardson2}, and \ref{fig:vort}. Since the term $\tau_a$ can advect vorticity but not create it, such vortices can only be created by accumulating flows which already have $\sgn{\omega_\phi}=\sgn{z}$. This process can be seen in Fig. \ref{fig:KHsimRichardson2}, where it is shown that the shear layer (with $\sgn{\omega_\phi}=-\sgn{z}$) is adjacent to a region of opposite $\omega_\phi$-sign (see also Fig. \ref{fig:vort_generation}). Once the KHI is triggered\footnote{Although the KHI is first triggered in the layer with $\sgn{\omega_\phi}=-\sgn{z}$, Fig. \ref{fig:KHsimRichardson2} shows that the adjacent layer with the opposite vorticity should also be KH-unstable (typically with a smaller growth rate), which could also facilitate the formation of eddies with $\sgn{\omega_\phi}=\sgn{z}$.}, such regions form vortices with $\sgn{\omega_\phi}=\sgn{z}$. Eddies created in this way can then be accelerated in the constant-$j_z$ zones by $\tau_b$.

In the vorticity distribution for $f_\mathrm{dg}=10^{-4}$ (Fig. \ref{fig:vort}), it can be seen that the amplified eddies are always adjacent to the zones with $\sgn{\omega_\phi}=\sgn{z}$ located at the smaller-$r$ side of the shear layers (it is harder to see this for $f_\mathrm{dg}=10^{-3}$, in which case the flow structure is less orderly). It is then possible that these vortices are being constantly fed gas with their same vorticity sign, possibly contributing to their acceleration. This seems to be a minor effect in the irradiated upper layers, where only the vortices with $\sgn{\omega_\phi}=\sgn{z}$ prevail, while vortices with $\sgn{\omega_\phi}=-\sgn{z}$ are inhibited by $\tau_b$. Close to the midplane, instead, the baroclinic torquing is weaker (Fig. \ref{fig:vort_generation}), and eddies rotate with $\sgn{\omega_\phi}=-\sgn{z}$ as determined by the KHI driving, while only a few lower-amplitude counter-rotating eddies survive.

In summary, the observed vortices with $\sgn{\omega_\phi}=\sgn{\tau_b}$ accelerated in the uniform-$j_z$ bands are seeded at the band interfaces via $\tau_a$ by the nonlinear evolution of the KHI\footnote{It is also possible that some vortices are spontaneously generated inside of the bands, although we have not verified this.}. The amplified vortices are rapidly destroyed when advected into the shear layers (possibly with some contribution of $\tau_c$), and thus their size is limited by the width of the uniform-$j_z$ regions.

 The prevailing baroclinic torque at the constant-$j_z$ bands may also explain why the VSI-induced large-scale flows also respect a rotation pattern with $\sgn{\omega_\phi}=\sgn{\tau_b}$. In its linear stage, the VSI transports gas across constant-$j_z$ surfaces, as determined by the unstable directions of motion predicted by linear theory \citep[][]{JamesKahn1970,Knobloch1982,Klahr2023a}. Since $\nabla j_z$ decreases with height and increases with radius, this transport reduces the $j_z$ gradient between gas parcels moving away from the midplane and their adjacent outer parcels moving toward it, which starts forming constant-$j_z$ bands. Upward- and downward-directed flows are eventually connected at the upper disk layers, either due to the angular momentum excess of the parcels moving away from the midplane or due to the rotation produced by the baroclinic torque once $|\tau_b|>|\tau_c|$. This rotation further enhances the angular momentum exchange between the connected regions, until bands of uniform $j_z$ are formed, inside of which $|\tau_b|\gg|\tau_c|$. Thus, the baroclinic acceleration may explain why the saturated VSI modes survive despite the lowered angular momentum gradient: even though the VSI driving is initially of centrifugal origin, it is the baroclinic torque what drives this rotation pattern once the centrifugal acceleration gradient vanishes.

\subsection{A closer look at meridional vortices}\label{SS:MeridionalVorticesAmpl}

\begin{figure}[t!]
\centering
\includegraphics[width=\linewidth]{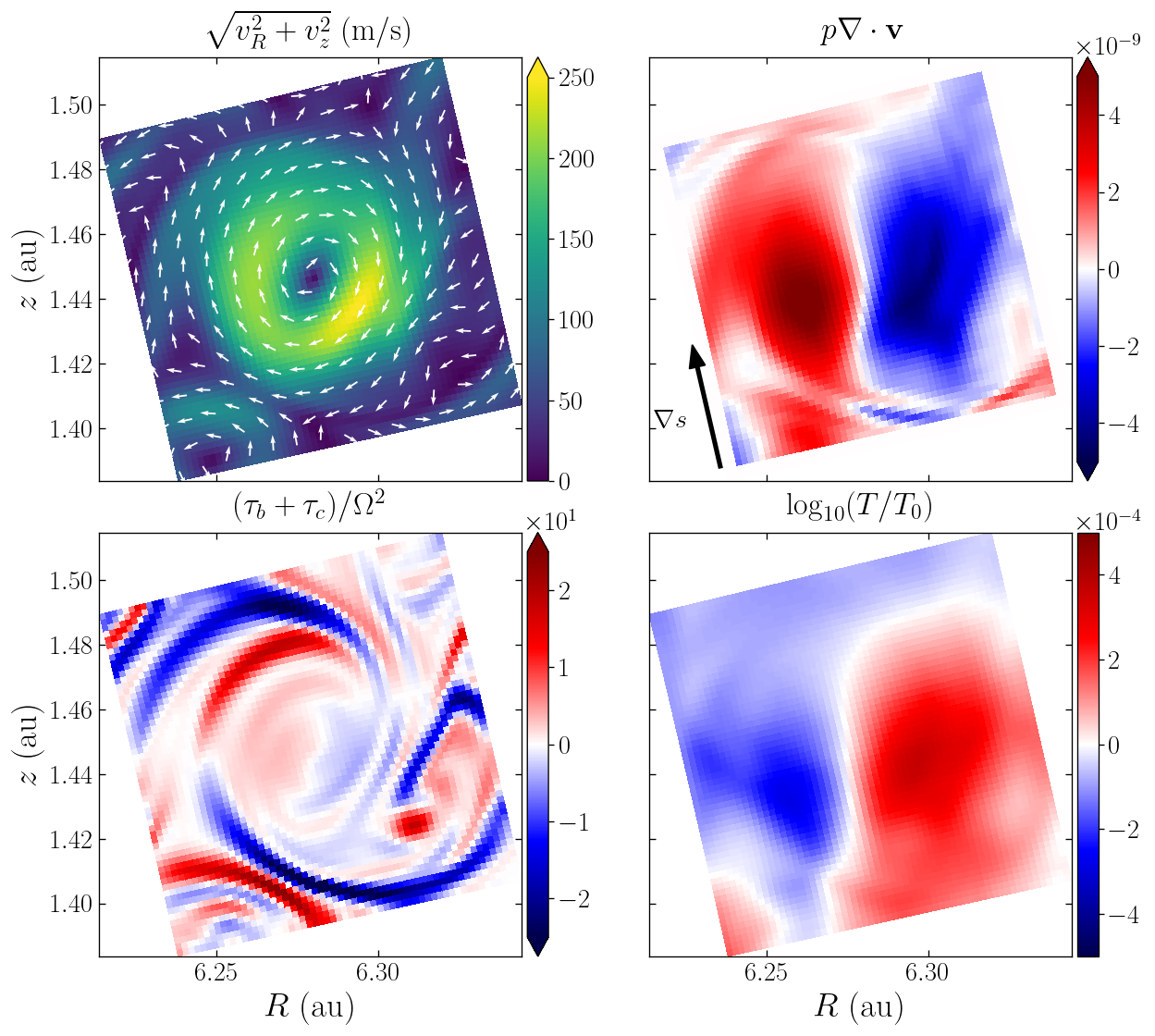}
\caption{ Vortex highlighted in the upper left panel of Fig. \ref{fig:vort}. Clockwise from the top left: meridional velocity norm, expansion work term (code units), temperature perturbation, and sum of the baroclinic and centrifugal torque terms in Equation \eqref{Eq:VorticityIntegral} normalized by $\Omega^2$. The direction of the meridional velocity and the entropy gradient are indicated with arrows in the left and right top panels, respectively.}
\label{fig:vortex}
\end{figure}

A gas parcel embedded in a vortex travels through regions of varying temperature, and thus experiences successive cooling and heating cycles. If the thermal relaxation time was longer than the eddy overturn time, the vortex temperature would become approximately uniform, which would in turn get rid of the local baroclinicity, since $\tau_b\propto \nabla \rho\times\nabla T$ (Equation 5 in Paper I). Therefore, in order to work, this process requires shorter cooling times than the rotation period of the vortices. This condition is satisfied in the entire domain, as the vortex rotation angular frequency, $\Omega_v$, is typically on the order of the orbital frequency, while the cooling time above the irradiation surface, where vortices are formed, is $10^{-4}-10^{-2}\,\Omega^{-1}$ for $f_\mathrm{dg}=10^{-3}$ and $10$ times larger for $f_\mathrm{dg}=10^{-4}$ (Section \ref{SS:VSIcooling}). On the other hand, despite these short cooling times, we can see some heating and cooling throughout the vortices, as seen in the temperature perturbations in Fig. \ref{fig:vortex}. This occurs due to the expansion and contraction, evidenced by the sign of the work term $p\nabla\cdot\mathbf{v}$ in that figure, caused by thermal relaxation as gas is transported to and from regions of increasing entropy, respectively (the direction of the entropy gradient is indicated on the same figure). The integral of this term in a region occupied by an amplified vortex is positive, indicating the conversion of internal to kinetic energy expected from the acceleration mechanism. 


An explanation for the fact that $\Omega_v=\mathcal{O}(\Omega)$ can be given by noting that bands of approximately constant angular momentum are radially compressed by the imbalance between gravity and centrifugal acceleration \citep[see, e.g.,][]{PapaloizouPringle1984}. Neglecting  $\mathcal{O}(z/R)^2$ terms, the sum of these  accelerations is $j_z^2/R^3-G M_s/R^2$, which is negative for large radii and positive for small radii. If one such band is radially kept in place by the pressure of its adjacent regions, this quantity must be be positive at the inner edge and negative at the outer edge of the band. In that case, neglecting meridional velocity fluctuations, hydrostatic balance in the radial direction determines that $\frac{1}{\rho}\partial_R p \approx j_z^2/R^3-G M_s/R^2$, which means that the pressure must reach a maximum at some radius $R=R_0$ inside of the band. This explains the positive pressure perturbations inside of the constant-$j_z$ bands in Fig. \ref{fig:vort}. Assuming that relative temperature variations are much smaller than relative pressure variations, which can be done as long as the cooling time is much smaller than the dynamical time ($t_\mathrm{cool}\ll\Omega^{-1}$), we can solve the resulting equation for $p$ by writing $\rho=p/c_s^2$ with constant $c_s^2$, which for small radial departures $\Delta R\ll R_0$ results in $\Delta p/p=|p(R_0+\Delta R)-p(R_0)|/p(R_0)\approx \frac{\Omega_0^2 \Delta R^2}{2 c_s^2}$, where $\Omega_0=\Omega(R_0)$. On the other hand, the vortices produce pressure minima that compensate for the outward centrifugal force due to their own rotation, as it can be seen in Fig. \ref{fig:vort}. For a vortex of radius $\sim \Delta R$, taking $\Delta R$ as the half-width of a VSI mode, this balance results in a relative pressure decrease $\Delta p/p\sim \frac{\Omega_v^2 \Delta R^2}{c_s^2}$. Thus, vortices can only be accelerated as long as they can produce deeper pressure minima, but this process is limited by the pressure increase caused by the continuously enforced radial contraction of the constant-$j_z$ bands. We can then expect the saturation of the vortex amplification to be reached when the pressure perturbation caused by both processes is of a similar order of magnitude, which leads to $\Delta p/p\sim \frac{\Omega_v^2 \Delta R^2}{c_s^2} \sim \frac{\Omega_0^2 \Delta R^2}{2 c_s^2}$, and so $\Omega_v\sim \Omega$. 

In their saturated state, the amplified eddies alter the fluid state in such a way that $\tau_c$ is no longer zero but of the same order of magnitude as $\tau_b$, which is also altered with respect to the initial state. Each of these terms individually have opposite signs in different regions of the saturated amplified vortices, and so does their sum, which controls the vortex's acceleration (Equation \eqref{Eq:VorticityIntegral}). Despite the generally complex form of this quantity (see Fig. \ref{fig:vortex}), we typically see that the negative regions approximately balance the positive regions when integrated inside of closed streamlines. This behavior is expected, as such a balance is required in order to halt the acceleration process.





%% file: longtime.tex
\subsection{Long-time evolution}\label{A:LongTime}

\begin{figure}[t]
\centering
\includegraphics[width=\linewidth]{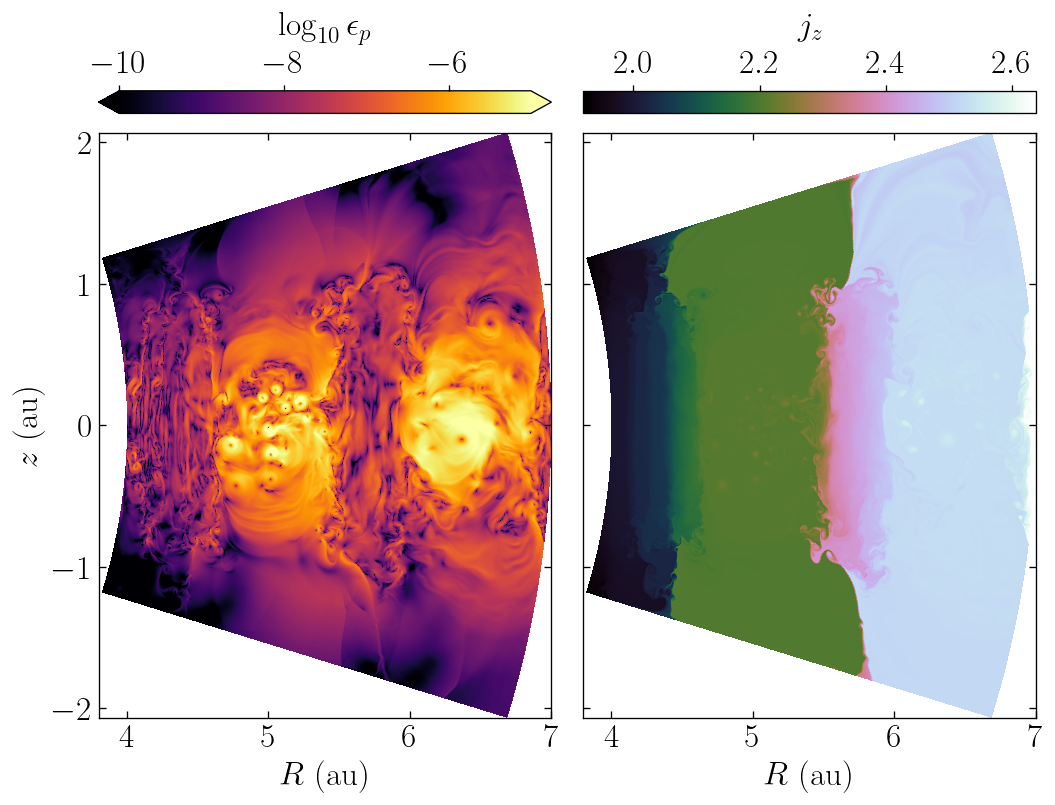}
\caption{Kinetic energy density due to meridional motion ($\epsilon_p=\frac{\rho}{2}(v_R^2+v_z^2)$) (left) and specific angular momentum (right) after 1360 orbits in run \sftw{dg3c4\_1024}.}
\label{fig:ekjzlongtime}
\end{figure}

In axisymmetric VSI simulations, bands of approximately constant $j_z$ can grow up to large widths ($\sim 1$ au in our simulations) for long enough times. This occurs due to the mixing of angular momentum enforced to occur in the meridional plane, given that these structures are unaffected by non-axisymmetric modes that should prevent them from growing to such extents in 3D. Since the size of the baroclinically amplified eddies is only limited by the width of the bands, these eddies are allowed to occupy larger regions as the bands grow. For $f_\mathrm{dg}=10^{-3}$, in which case the VSI is active in the entire domain, the $j_z$ gradient is eventually largely reduced in most of the domain (except at the few remaining interfaces between bands) after thousands of orbits. In that state, most of the kinetic energy is contained in the amplified vortices, as shown in Fig. \ref{fig:ekjzlongtime}. Although we do see vortex mergers, potentially indicating an inverse energy cascade resulting from the enforced axisymmetry, it is unclear whether it is this phenomenon or the baroclinic driving that is the main mechanism behind the formation of the large eddies (see Section \ref{SS:EnergyTransportScales}).

\begin{figure}[t]
\centering
\includegraphics[width=\linewidth]{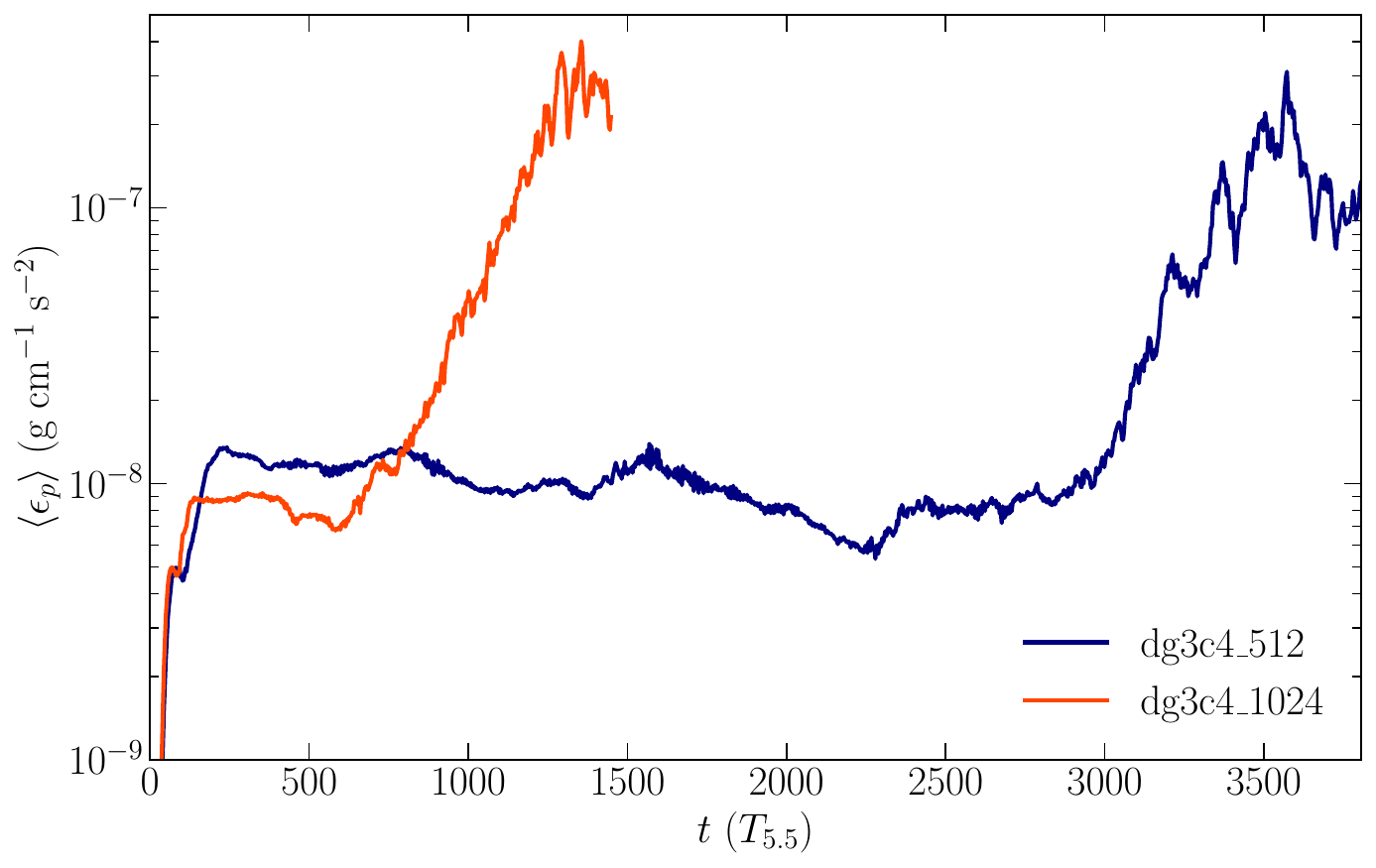}
\caption{Time evolution of the average meridional kinetic energy in runs \sftw{dg3c4\_512} and \sftw{dg3c4\_1024}.}
\label{fig:ekintlongtime}
\end{figure}

Even if the resolution is not large enough to initially resolve the amplified eddies, as it occurs for $N_\theta=512$, these are eventually resolved for sufficiently long times due to the increase of their typical size resulting from the growth of the bands. The transition into this regime occurs at earlier times for increasing resolution, as evidenced by the time evolution of the kinetic energy (Fig. \ref{fig:ekintlongtime}). This shows that the mixing of angular momentum in between constant-$j_z$ regions is more efficient when smaller scales are resolved, and in particular when the shear in between these regions is larger, which may result from the angular momentum mixing between adjacent bands produced by the KH eddies, evidenced, for instance, in Fig. \ref{fig:veljz}. Moreover, this also shows that the observed transition is not caused by numerical diffusion of angular momentum, as that would lead to the opposite trend.


A transition into an eddy-dominated regime is also seen in isothermal versions of our setup (not shown here) and in the isothermal runs by \cite{Klahr2023b}. This shows that, even though the increased optical depth of the large eddies with respect to the VSI modes and the subsequent increase of the local cooling timescales in the Rad-HD simulations could potentially trigger the COS, that instability is not what produces the transition into that regime. The required high resolutions and long times to obtain this transition (in our case $\sim 3000$ and $800$ orbits for $50$ and $100$ cells per scale height, respectively, with our employed
numerical scheme) explain why this phenomenon has not been seen in other works. 

The mentioned expected differences with non-axisymmetric simulations suggest that this transition is merely a consequence of the enforced axisymmetry. Thus, the long-term nonlinear evolution of the VSI can only be studied in high-resolution 3D simulations (see Sections \ref{SS:EnergyTransportScales} and \ref{SS:AxisymmetricCaveats}).

%% file: stability1.tex
\section{Stability regions}\label{S:StabilityAnalysis}

\subsection{Influence of collisional timescales and gas emissivity}\label{SS:VSIcooling}

In Paper I, we analyzed the stability of our Rad-HD simulations in terms of the local criterion by \cite{Urpin2003}, which predicts instability as long as
\begin{equation}\label{Eq:tcrit_loc}
t_\mathrm{rel}\lesssim\frac{|\partial_z v_\phi|}{N_z^2}\,,
\end{equation}
where $t_\mathrm{rel}$ is the local thermal relaxation time, which in the Rad-HD simulations coincides with the radiative cooling time $t_\mathrm{cool}$, while $N_z$ is the vertical Brunt-V\"ais\"al\"a frequency \citep[e.g.,][]{Rudiger2002}. The localized suppression of the VSI in the middle layer of our dust-depleted disk is then explained by the fact that this criterion is not met in that entire region. On the other hand, the vertically global stability criterion by \cite{LinYoudin2015}, predicting instability for
\begin{equation}\label{Eq:tcrit_glob}
  t_\mathrm{rel}<\frac{|q|}{\Gamma-1}
  \frac{H}{R}\,
  \Omega_K^{-1}\,,
\end{equation}
where $\Gamma$ is the heat capacity ratio, $q=\frac{\partial \log T}{\partial \log R}$, and $\Omega_K$ is the Keplerian rotation frequency, correctly predicts the stability properties of the middle layer but does not contemplate cases in which the disk surface layers, where the assumptions of vertically uniform temperature and cooling timescale break down, have different stability conditions than the midplane.

The stability analysis in Paper I relies on the assumption that gas and dust particles are in local thermal equilibrium. However, this is not the case at the surface layers, whose low densities make collisions between dust and gas particles infrequent enough that these species may have different temperatures. Considering this, we can still estimate $t_\mathrm{rel}$ in those regions by means of a linear analysis of small perturbations of separate dust and gas temperatures. We follow this approach in Appendix \ref{A:CollisionalTimescale} in the same way as \cite{Barranco2018}, with the difference that our largest considered dust grains are micron-sized, whereas in that work cm- to meter-sized grains were assumed. Despite this difference, the functional form of the obtained timescale is identical as in that work\footnote{As mentioned in Paper I, a factor of $1/\Gamma$ is missing in the definition of the thermal relaxation timescale, since we have omitted the expansion work terms in our timescale computation. However, this factor can be safely omitted for the purpose of a stability analysis.}:
\begin{equation}\label{Eq:trel_ms}
    t_\mathrm{rel}=t_g + t_\mathrm{cool}\,,
\end{equation}
where the radiative cooling timescale, $t_\mathrm{cool}$, is computed as in Paper I, while the collisional characteristic timescale for gas thermal relaxation, $t_g$, is obtained as
\begin{equation}
    t_g=\frac{4}{3(\Gamma-1)}\left(\frac{\rho_{gr}}{\rho^\mathrm{tot}_d}\right)\frac{(a_\mathrm{max}a_\mathrm{min})^{1/2}}{\overline{v_g}}\,,
\end{equation}
where $\rho_\mathrm{gr}$ is the bulk density of the dust grains, 
$\rho^\mathrm{tot}_d$ is the total local dust density, $a_\mathrm{max}$ and $a_\mathrm{min}$ are respectively the maximum and minimum grain radii, and $\overline{v_g}=\sqrt{\frac{8 k_B T}{\pi \mu u}}$ is the mean thermal speed of gas particles. For an assumed total dust-to-gas ratio  $\rho^\mathrm{tot}_d/\rho=10^{-2}$, our assumed dust-to-gas ratios for small $<0.25$ $\mu$m grains, $f_\mathrm{dg}=10^{-3}$ and $10^{-4}$, correspond respectively to $a_\mathrm{max}=19$ $\mu$m and $1.8$ mm.
Even if dust settling imposes a lower cutoff to the maximum dust size than $a_\mathrm{max}$, the value of $t_g$ is unchanged, as shown in Appendix \ref{A:CollisionalTimescale}. We also show in the same appendix that our obtained expression for $t_\mathrm{rel}$ is unchanged if the dust temperature depends slowly on the grain size.

\begin{figure}[t!]
\centering
\includegraphics[width=\linewidth]{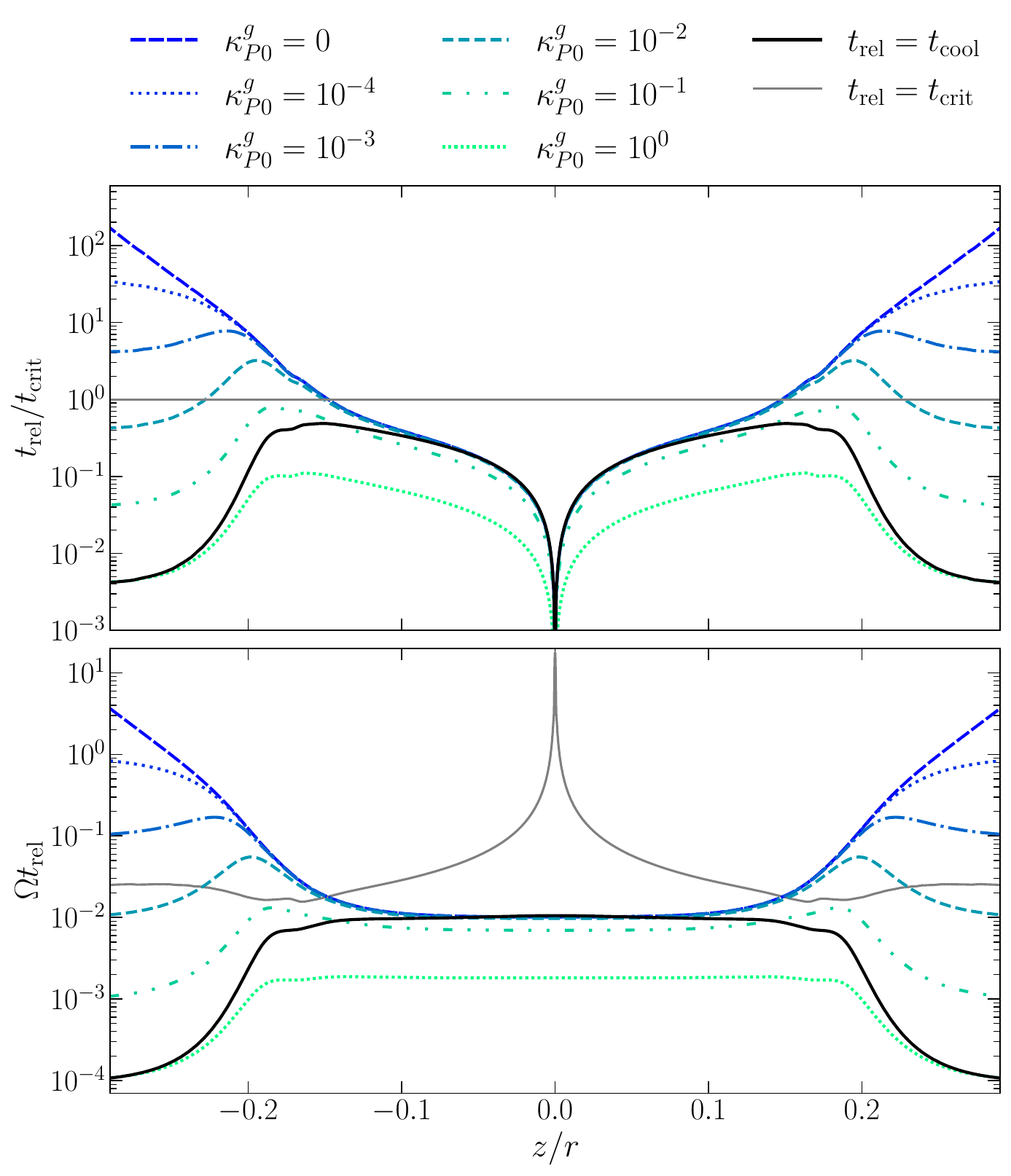}
\caption{Thermal relaxation times normalized by the critical cooling time $t_\mathrm{crit}$ (top) and the orbital frequency (bottom) at $r=5.5$ au for $f_\mathrm{dg}=10^{-3}$, assuming only radiative cooling ($t_\mathrm{rel}=t_\mathrm{cool}$), including the effect of dust collisions ($\kappa^g_{P0}=0$), and including gas emission with $\kappa^g_{P0}$ defined in Equation \eqref{Eq:GasOpac}.
Also shown are the corresponding values for $t_\mathrm{rel}=t_\mathrm{crit}$. The gas opacities reported in \cite{Freedman2008}, \cite{Malygin2014}, and \cite{Malygin2017} for solar abundances correspond to $\kappa^g_{P0}\sim 10^{-1}-10^{0}$.
}
\label{fig:tcool_tcrit_dg3}
\end{figure}

\begin{figure}[t!]
\centering
\includegraphics[width=\linewidth]{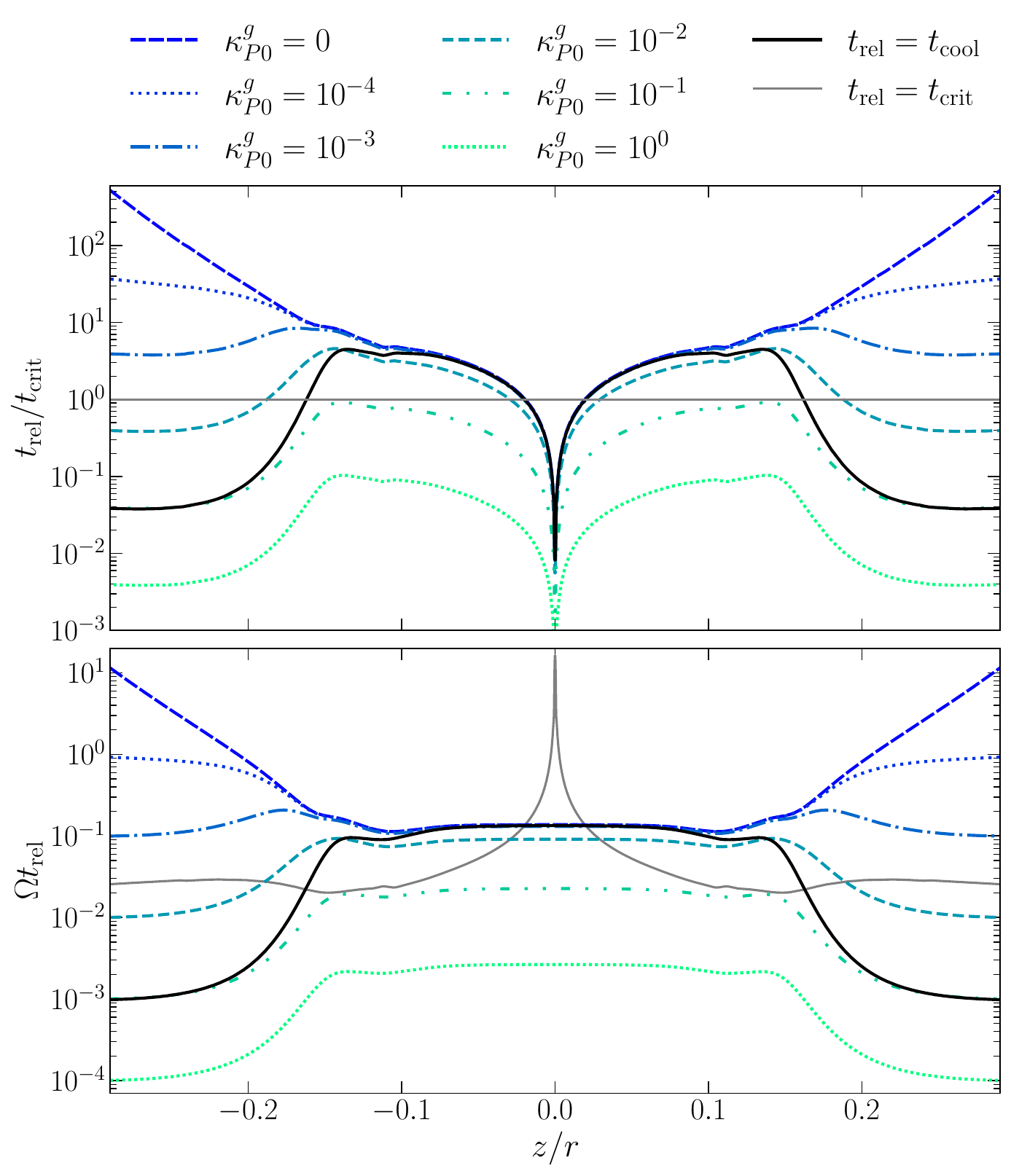}
\caption{Same as Figure \ref{fig:tcool_tcrit_dg3} but for $f_\mathrm{dg}=10^{-4}$.}
\label{fig:tcool_tcrit_dg4}
\end{figure}

The thermal relaxation times in our simulations, computed using Equation \eqref{Eq:trel_ms}, are shown in Figures \ref{fig:tcool_tcrit_dg3} and \ref{fig:tcool_tcrit_dg4} (labeled as $\kappa_P^g=0$), where it can be seen that the increase of the thermal relaxation time at the surface layers can inhibit the growth of VSI modes, as it occurs, for example, in the hydrodynamical simulations by \cite{Pfeil2021}. For $f_\mathrm{dg}=10^{-3}$, the VSI can still operate up to a few scale heights above the midplane, while the instability should be nearly completely suppressed for $f_\mathrm{dg}=10^{-4}$.

This situation changes if cooling by molecular emission is not negligible. This effect should be most important at the surface layers, where the typical temperatures of up to $\sim 200-250$ K in our domain may increase the amount of possible free-free and bound-free emission (and absorption) mechanisms \citep[e.g.,][]{Freedman2008,Malygin2014} with respect to the cooler midplane. If optically thin radiative emission by gas molecules is considered, it can be shown (Appendix \ref{A:CollisionalTimescale}) that the linear relaxation timescale is modified as
\begin{equation}\label{Eq:trelkPg_ms}
    t_\mathrm{rel} = t_\mathrm{rel,thin}  \frac{3+\lambda^{dg}_R\lambda^{dg}_P k^2}{\lambda^{dg}_R\lambda^{dg}_P k^2}\,,
\end{equation}
where $t_\mathrm{rel,thin}$ is the optically thin limit of this expression, given by
\begin{equation}\label{Eq:trelkPg_thin_ms}
    t_\mathrm{rel,thin} =
    \left(
    \frac{1}{t_g+t_\mathrm{cool,thin}}
    +
    \frac{1}{t_{r,g}}
    \right)^{-1}\,,
\end{equation}
where $\lambda^{dg}_{P}$ and $\lambda^{dg}_{R}$ are respectively the Planck and Rosseland mean free paths due to both gas and dust photon absorption (see Appendix \ref{A:CollisionalTimescale}), while $t_\mathrm{cool,thin}$ is the optically thin limit of $t_\mathrm{cool}$ (Eq. 17 in Paper I) and $t_{r,g}$ is the optically thin cooling time due to gas emission. Assuming that the gas emission rate is determined by Kirchhoff's law, we compute the latter timescale as $t_{r,g}=\frac{c_g T}{c \kappa_P^g [4a_RT^4+b^g_P(a_RT^4-E_r)]}$, where $c_g=\frac{k_B}{(\Gamma-1)\mu u}$ is the gas specific heat capacity at constant volume, $\kappa_P^g(T)$ is the Planck-averaged gas absorption opacity, and $b^g_P=\frac{d \log \kappa_P^g}{d\log T}$. Gas opacities can in general span several orders of magnitude ($\sim 10^{-6}-1$ cm$^2$ g$^{-1}$) depending on chemical composition and molecular abundances \citep{Freedman2008,Helling2009,Dullemond2010,Malygin2014}. If we assume, as we have done throughout this work, that the temperatures of the gas and the reprocessed radiation are similar, then we should also expect the gas opacity to drop at low temperatures \citep[$\lesssim 75 K$ in][assuming solar composition]{Freedman2008,Malygin2014} due to the lack of absorption lines at low frequencies. Taking these considerations into account, we adopt a simplified gas opacity law of the form
\begin{equation}\label{Eq:GasOpac}
    \kappa^g_P(T)=\kappa^g_{P0}  \,
    \Theta(T-75\,\mathrm{K})\,
    \mathrm{cm}^{2}\,\mathrm{g}^{-1}\,,
\end{equation}
where $\Theta$ is the Heaviside function and the constant $\kappa^g_{P0}\in[10^{-4},1]$ parameterizes the uncertainty in the gas composition. For the computation of the combined dust-gas Rosseland opacity, we only consider the contribution of dust absorption, assuming that our densities are low enough that absorption lines can be disregarded in this calculation \citep[see, e.g.,][]{Freedman2008}.

The obtained relaxation times at $5.5$ au for different $\kappa^g_{P0}$ values are shown in Figures \ref{fig:tcool_tcrit_dg3} and \ref{fig:tcool_tcrit_dg4}. These figures show that the VSI can still operate at the surface layers provided gas opacities are large enough ($\kappa^g_{P0}\gtrsim 10^{-2}$ in our models). For $\kappa^g_{P0}\gtrsim 10^{-1}$, the gas opacities are large enough that even the middle layer is VSI-unstable in our dust-depleted case. However, we can only expect the temperature and density distributions to be roughly unchanged for varying opacity if $\kappa^g_P$ is smaller than both $f_\mathrm{dg}\kappa^d_P$ and $f_\mathrm{dg}\kappa^d_R$, where $\kappa^d_P$ and $\kappa^d_R$ are respectively the Planck and Rosseland means of the dust absorption opacity.
Using these criteria, we estimate the computed $t_\mathrm{rel}/t_\mathrm{crit}$ values in our disk models to be reliable at the middle layers for $\kappa^g_{P0}\lesssim 10^{-1}$ and $10^{-2}$ for $f_\mathrm{dg}=10^{-3}$ and $10^{-4}$, respectively, while at the upper layers the temperatures the temperature and density distributions should be unchanged for $\kappa^g_{P0}\lesssim 10^{-1}$. Thus, our conclusion regarding the possible instability of the upper layers remains the same.


%% file: stability2.tex
\begin{figure*}[t!]
\centering
\includegraphics[width=\linewidth]{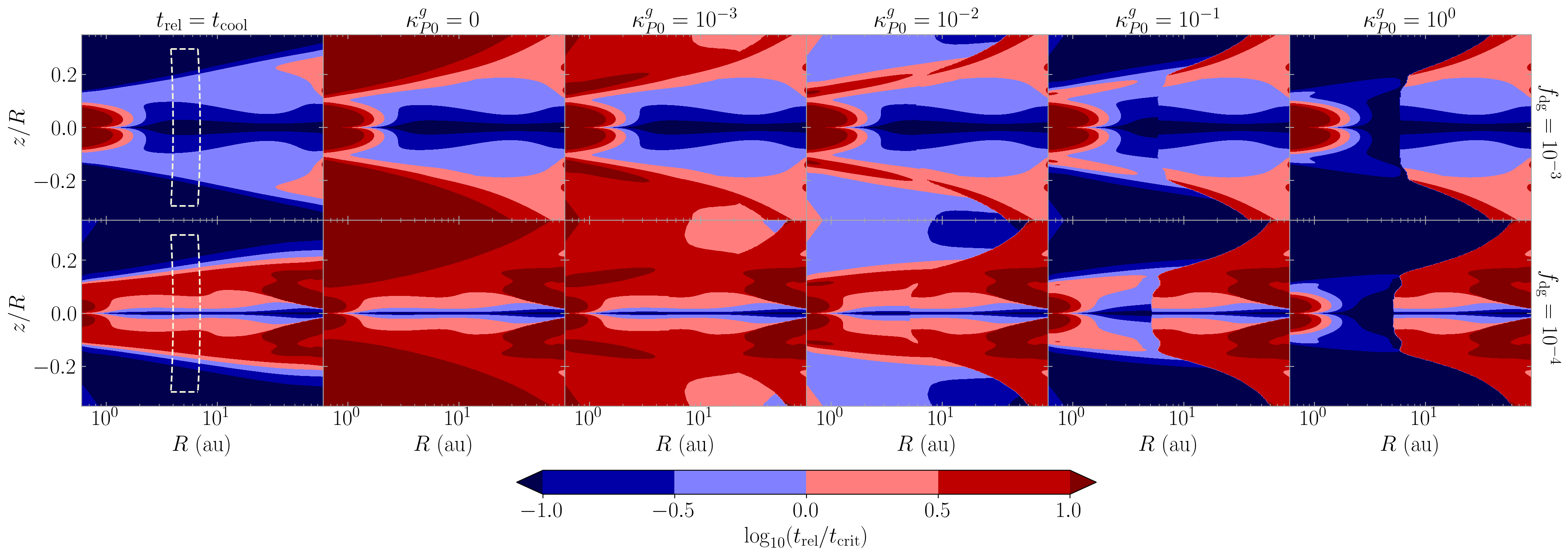}
\caption{VSI-stability maps showing the ratio of the relaxation time to the critical cooling time defined in the local stability criterion (Eq. \eqref{Eq:tcrit_loc}). From left to right, $t_\mathrm{rel}/t_\mathrm{crit}$ values are computed assuming only radiative cooling ($t_\mathrm{rel}=t_\mathrm{cool}$), including dust collisions ($\kappa^g_{P0}=0$), and including as well molecular emission with $\kappa^g_{P0}\neq 0$ (Equation \eqref{Eq:GasOpac}). The top and bottom row correspond to the nominal and dust-depleted cases, respectively. Also shown in the leftmost panels is the location of the domain used in our Rad-HD simulations (white dashed box).
 The gas opacities reported in \cite{Freedman2008}, \cite{Malygin2014}, and \cite{Malygin2017} for solar abundances correspond to $\kappa^g_{P0}\sim 10^{-1}-10^0$.
}
\label{fig:tctcr_fulldomain_2D}
\end{figure*}

\begin{figure*}[t!]
\centering
\includegraphics[width=\linewidth]{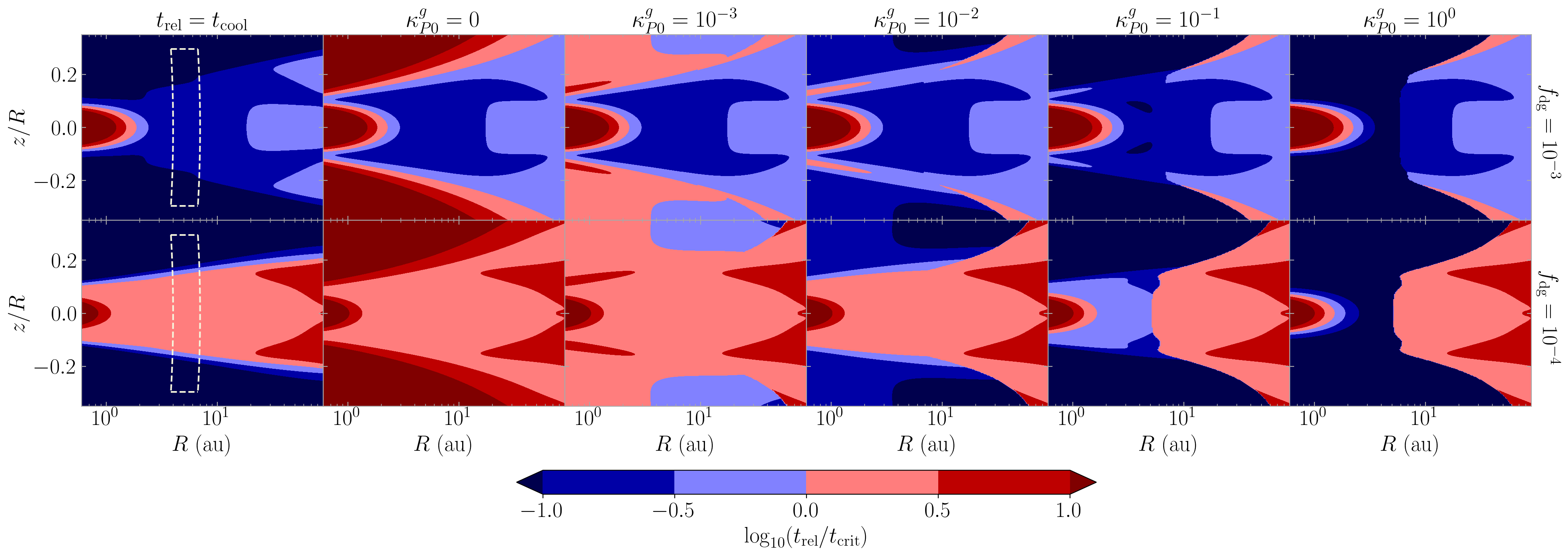}
\caption{Same as Figure \ref{fig:tctcr_fulldomain_2D} but locally computing $t_\mathrm{crit}$ as in the right-hand side of the global stability criterion (Equation \eqref{Eq:tcrit_glob}).}
\label{fig:tctcr_fulldomain_2D_globalcrit}
\end{figure*}

\subsection{Global disk stability}\label{SS:GlobalDiskStability}

In this section we extend our stability analysis to the full domain of the hydrostatic disk models used as initial conditions (see Paper I), for which an estimation of the dominant VSI wavenumbers is required. To compute $t_\mathrm{cool}$, we assume the maximum dominant wavenumbers in our simulations to be approximately $k\sim 70/H$ (Fig. C.3 in Paper I) in all disk regions. For larger radii than in our simulation domain, this does not introduce any error, given that our obtained VSI modes are radially optically thin (Paper I) and therefore cool at the scale-independent optically thin rate given by Equation 16 of that work, while for larger radii the optical depth along any wavelength only decreases with radius. For smaller radii, on the other hand, we can instead expect the dominant wavelengths in our simulations to become optically thick close to the midplane at some minimum radius, in which case VSI modes of such wavelengths cool via diffusion. If $t_\mathrm{rel}>t_\mathrm{crit}$ at some location for our considered $k$ value, that just means that only modes with larger wavenumbers can be unstable at that point. Considering that the strength of the instability decreases in our simulations for increasing average wavenumber (Paper I), it is also likely that this is also translated into either a partial or total suppression of the VSI.

The resulting $t_\mathrm{rel}/t_\mathrm{crit}$ values computed in this way are shown in Fig. \ref{fig:tctcr_fulldomain_2D} between $0.6$ and $90$ au (this is far enough from the radial boundaries to avoid boundary artifacts, as we used zero-gradient conditions at those boundaries to construct the hydrostatic models). Interpreting these distributions as approximate VSI stability maps, we obtain that the unstable surface regions extend up to a few tens of au for high enough gas opacity ($\kappa^g_{P0}\gtrsim 10^{-2}$). For $f_\mathrm{dg}=10^{-3}$, the midplane is unstable between $\sim 2$ au (VSI modes cool via diffusion for smaller radii), and $\sim 100$ au, after which densities are low enough that the optically thin cooling timescale is above $t_\mathrm{crit}$. For  $f_\mathrm{dg}=10^{-4}$, the entire midplane is stable unless $\kappa^g_{P0}\gtrsim 10^{-1}$, in which case an instability region is formed between $1$ and $5$ au due to the decrease of the optically thin cooling time with $\kappa^g_{P0}$.

For comparison with \cite{Malygin2017}, we also computed stability maps based on a local evaluation of Equation \ref{Eq:tcrit_glob} as done in that work, which we show in Figure \ref{fig:tctcr_fulldomain_2D_globalcrit}. Both criteria coincide in that region for the following reasons: (I) the assumptions of the global criterion are met below the irradiation surface, and (II) the vertically global critical cooling time (right-hand side of Equation \eqref{Eq:tcrit_glob}) can be obtained by evaluating $t_\mathrm{crit}$ at $z=\Gamma H/2$, which is much smaller than the vertical extent of the middle layer.
We also see a general agreement on the predicted unstable regions at the surface layers, with some differences for small $\kappa^g_{P0}$. We defer a thorough comparison of local stability criteria with hydrodynamical simulations to future works.

Differences with the stability maps at the midplane in \cite{Flock2017RadHydro,Flock2020} and \cite{Pfeil2019,Pfeil2021}, in particular the fact that an inner stable region and an outer stable region are obtained in the latter, are explained by the different assumed dust masses, opacity laws, and dominant wavelengths in those works and this one. The radiative cooling time is in general diffusion-dominated in inner regions and thus it increases with increasing opacity \citep[e.g.,][]{LinYoudin2015}, which explains the inner stable region in \cite{Pfeil2021} and this work. Conversely, far enough from the star that perturbations are optically thin, $t_\mathrm{cool}$ grows for increasing radius as the photon mean free path increases (see Paper I), eventually reaching the condition $t_\mathrm{cool}>t_\mathrm{crit}$, as shown in Fig. \ref{fig:tctcr_fulldomain_2D} (see also \cite{Malygin2017} and \cite{Dullemond2022}). The fact that the cooling times in \cite{Flock2017RadHydro,Flock2020} are estimated to be diffusion-dominated between $20$ and $100$ au for $f_\mathrm{dg}=10^{-3}$ despite the similar parameters in that model and ours is explained by the approximation made in those works that the dominant VSI wavelength is $\sim H$.

The most striking difference with other stability maps is the presence of unstable regions at the surface layers for reasonably high gas opacities, where we should expect a localized VSI activity as seen in our dust-depleted simulations. These regions appear as a consequence of the reduction of the thermal relaxation times above the irradiation surface induced by our temperature and density stratification, and thus they do not appear in stability maps of vertically isothermal disks \citep[][]{Malygin2017} nor in the models by \citep[][]{Pfeil2019}, in which the temperature is computed via an effective 1D treatment of vertical radiative diffusion. Since gas radiative emission can only reduce the relaxation timescale for large enough temperatures ($\gtrsim 100$ K is a rough estimate based on \cite{Malygin2017}), the radial extent of these unstable regions depends not only on the chemical composition of the gas but also on the dust opacity at the disk upper layers and the temperature of the central star, occupying larger radii for higher temperatures as evidenced by Equation 4 in Paper I. Future studies can test these results in hydrodynamical simulations by employing methods decoupling the dust and gas temperatures, such as that in \cite{Muley2023}.


%% file: discussion.tex
\section{Discussion}\label{S:Discussion}

\subsection{KHI in other works}

Although clearly identifying the KH eddies in our vorticity distributions requires resolutions of at least $N_\theta\geq 1024$, we also see similar velocity perturbations to the VSI modes as in Figs. \ref{fig:veljz} and \ref{fig:KHsim} for lower resolutions (Fig. A.2 in Paper I). This indicates that the KHI is also triggered in our lower-resolution runs ($N_\theta\geq 256$, $H/\Delta r \geq 25$) even though eddies are not well resolved. Similar signs of developing KHI can be seen in other works on the VSI showing velocity distributions resulting from both 2D and 3D simulations \citep[e.g.,][]{Nelson2013,Manger2018,Pfeil2021}, which indicates that this is a rather general phenomenon. This also suggests that the KHI is the instability mechanism behind the meridional vortices in the axisymmetric isothermal simulations by \cite{FloresRivera2020}.

\subsection{Saturation mechanism}\label{SS:SaturationMechanism}

The reduction of the energy proportion in large scales once the KHI is triggered (Section \ref{SS:SpectralAnalysis}) supports the hypothesis, proposed by \cite{LatterPapaloizou2018}, that the growth of the KHI plays an important role in the saturation of the VSI. In Paper I, it can be seen that the perturbations produced by the KHI to the VSI modes become more prominent for increasing resolution (Fig. A.2 in Paper I), as the shear between adjacent bands increases and the KH eddies are better resolved. Moreover, larger resolutions result in smaller average VSI wavelengths (Fig. C.3 in Paper I), and therefore also in a larger density of KH-unstable interfaces. All of these effects may contribute to the obtained reduction of the saturated average kinetic energy and Reynolds stresses with increasing resolution shown in Paper I.

The distribution of the gas velocity in relation to the location of the KH-unstable zones provides a rough picture of this saturation mechanism: once the VSI
modes reach sufficiently high velocities to trigger the KHI, the loss of energy to KH eddies in the unstable shear layers acts as an effective friction between the vertically moving flows (see Fig. \ref{fig:KHdiagram}), limiting the maximum kinetic energy that these can gain. Since in our simulations only a very small fraction of the energy is in the highest wavenumbers (Fig. \ref{fig:spectra}), we expect numerical dissipation to play a minor role in the total energy dissipation as long as a direct energy cascade does not occur (see Section \ref{SS:EnergyTransportScales}). In real 3D disks, viscous dissipation could become important if the KHI triggers a turbulent cascade leading to dissipation down to molecular viscous lengthscales. Instead, in our simulations, it is likely that energy is radiatively dissipated at the medium to large scales of the VSI flows and KH eddies. In particular, some dissipation may be produced by the baroclinic deceleration of the primary KH eddies, which grow with $\sgn{\omega_\phi}=-\tau_b$.


Another possible saturation mechanism is the parametric instability considered in \cite{CuiLatter2022}, whose longest wavelengths are expected to require $\sim 30$ grid points per VSI wavelength to be resolved. In our simulations, the wavelength of the VSI modes is well resolved with $\lambda/\Delta r\sim 15-75$, and in particular $\sim 75$ in Fig. \ref{fig:KHsim}. In that figure we do indeed see small-scale perturbations inside of the VSI modes, but we do not see significant amounts of kinetic energy being transferred to them (see also Fig. \ref{fig:vort}). Rather, the wavelength of the VSI modes grows over time due to the mixing of angular momentum between approximately uniform-$j_z$ bands (Paper I and Section \ref{A:LongTime}), and so even if the ripples inside of the VSI modes are produced via parametric instability, that process is not able to destroy such modes. On the other hand, it is unclear whether the observed ripples are instead sound waves emitted at the KH-unstable regions. Thus, we cannot conclude that a parametric instability occurs in our simulations due to the current lack of unequivocal signs of its occurrence in its nonlinear stage, but even if it does, its dynamical effect in our simulations appears to be marginal.

\subsection{Energy transport between scales}\label{SS:EnergyTransportScales}

The transfer of energy from VSI to KHI modes does not imply the occurrence of an energy flux from large to small scales as it occurs in 3D turbulent cascades. Instead, as first studied by \cite{Kraichnan1967}, 2D flows typically exhibit a split cascade transporting energy toward large scales and enstrophy toward small scales \citep[some experimental verifications of such a cascade can be found in][]{Bruneau2005,Kelley2011}. Thus, it is possible that a similar phenomenon occurs in our simulations. In this picture, energy is continuously injected in a broad range of wavelengths due to the geometry of the large-scale VSI flows (as we have verified) and transported toward small $k$ starting from the largest $k$ allowing for KHI growth (Section \ref{SS:SpectralAnalysis}). However, it must be noted that our simulations are not strictly two-dimensional. In particular the mechanisms of vortex stretching (Appendix \ref{A:VorticityGeneration}) and strain self-amplification\footnote{Assuming incompressibility, the source term producing strain self-amplification is proportional to the determinant of the strain rate tensor, $S=\frac{1}{2}(\nabla\mathbf{v}+(\nabla\mathbf{v})^\intercal)$, which even for $v_\phi=0$ is not necessarily zero for axisymmetric flows as long as $v_R\neq 0$.}, often linked with the development of direct energy cascades in 3D \citep[see, e.g.,][]{Carbone2020,Johnson2021}, are absent in 2D flows but not in axisymmetric flows. Moreover, as reviewed by \cite{Alexakis2018} \citep[see also the discussion in][]{Sengupta2023}, both direct and inverse energy cascades are possible in 3D (non-axisymmetric) turbulent flows when the effects of rotation, stratification, and even compressibility become important.

In our simulations, a hint possibly indicating transport of energy toward lower wavenumbers is given by the observation that vortices merge and grow in size, likely contributing to the formation of large-scale meridional vortices at long times (Section \ref{A:LongTime}). Also resulting in the accumulation of energy in increasingly large scales is the growth over time of the uniform-$j_z$ bands, likely via in-plane angular momentum mixing between adjacent bands. If an inverse energy cascade occurs, then energy should be radiatively dissipated at the medium and large scales. In this scenario, it is possible that the slopes measured for $5\lesssim kH < 100-200$ result from a combination of an inverse cascade and the geometry imposed by the VSI modes. In this regard, we must clarify that, despite the spectra decay close to $k^{-5/3}$ for $f_\mathrm{dg}=10^{-3}$ as one would expect in fully developed homogeneous and isotropic turbulence (even in 2D), the flows in our simulations depart from these conditions, even in the $k$-range dominated by KHI modes. At such scales, anisotropy can still be observed even for the velocity components perpendicular to the VSI flows, which track mostly KHI modes. Moreover, the spectra deviate from that functional form after several tens of orbits, as the constant-$j_z$ regions grow. This leads to a steepening over time,  with $E(k)$ eventually oscillating in that wavenumber range around $\sim k^{-2.4\pm 0.2}$ for $f_\mathrm{dg}=10^{-3}$ and $\sim k^{-3\pm 0.2}$ for $f_\mathrm{dg}=10^{-4}$.

It is unclear whether the accumulation of energy at increasingly large scales is necessarily the result of an inverse cascade. Moreover, neither the observation of sporadic vortex mergers and the growth of uniform-$j_z$ regions, nor the evaluation of the slopes of the energy spectra, are sufficient to  estimate the influence of rotation, gravity, weak compressibility, and baroclinicity on the transport of energy and enstrophy between scales. Some information on this can likely be obtained by computing the energy and enstrophy fluxes resulting from each of these phenomena, as done in the multiple analyses in \cite{Alexakis2018} for incompressible flows. We defer such an analysis to a future work on the topic.

\subsection{Caveats of axisymmetric models}\label{SS:AxisymmetricCaveats}

Once the KHI is triggered in our simulations, amplified eddies never cease to form and dissipate, and thus they are able to continuously limit the maximum kinetic energy of the vertical flows even hundreds of orbits after saturation. Unfortunately, our simulations are insufficient to make predictions on the disk's state after thousands of orbits, given that the artificially enhanced growth of uniform-$j_z$ bands due to vertical angular momentum mixing in 2D eventually leads in some cases to a complete breakdown of the VSI flows and domination of large-scale azimuthal vortices, as shown in Section \ref{A:LongTime} \citep[see also][]{Klahr2023b}. If the growth of these vortices is enhanced via an inverse energy cascade, this process should likely be inhibited in 3D simulations, provided that a direct energy cascade occurs in that case. With enough azimuthal resolution, we expect the formation of wide uniform-$j_z$ bands \citep[Paper I and][]{PapaloizouPringle1984} to be inhibited by the growth of non-axisymmetric modes, which should disallow the growth of large-scale vortices at long times.

On top of this, it is even unclear whether small-scale eddies can be baroclinically amplified and survive for several orbits if axisymmetry is not imposed. In principle, all of the ingredients needed for the formation of meridional vortices could still be present in that case, given that the VSI driving mechanism is axisymmetric. As a result, the large-scale VSI-induced flows maintain some degree of axisymmetry in some 3D simulations \citep[e.g.,][]{Richard2016,Flock2020,BarrazaAlfaro2021} and are likely prone to KHI, as the velocity profiles in \cite{Manger2018} (Figure 10) seem to indicate. Axisymmetry, however, could be artificially enhanced by underresolving the $\phi$-direction, and thus it is in general unclear to which extent non-axisymmetric modes can erase the initial axisymmetry of the VSI modes. Therefore, the question remains of whether bands of reduced $\nabla j_z$ are formed in 3D, and even whether meridional vortices are destroyed by non-axisymmetric modes. As discussed in Paper I and later in this section, such regions appear to form in the 3D HD simulations by \cite{Richard2016}, \cite{Manger2018}, and  \cite{Flock2020} (see also Fig. A.1 in Paper I), but asserting the universality of this phenomenon requires further careful investigation.

Determining whether a baroclinic amplification of KH eddies can occur in non-axisymmetric disks has the additional difficulty that cell aspect ratios of $\sim1$ should be achieved to avoid purely axisymmetric effects, together with the high resolutions of at least $\sim 100$ cells per scale height needed to resolve the vortices, which is computationally rather expensive. Interestingly, radial resolutions of $H/\Delta R \sim 100-150$ are reached in the $\beta$-cooling 3D disk simulations by \cite{Richard2016} for an aspect ratio of $H/R\sim0.05$ similar to those in this work (see Paper I), albeit with a steeper temperature profile ($T\sim R^{-1}$ vs. $T\sim R^{-1/2}$ in this work) and with lower azimuthal and meridional resolutions than in the radial direction ($H/(R \Delta \phi) \approx H/(R \Delta \theta) \approx 20$). In those simulations, the VSI initially grows axisymmetrically, forming bands of reduced vertical vorticity $\omega_z$ with respect to the initial Keplerian state. These bands are separated by shear layers in which $\omega_z$ abruptly increases. When the VSI flows reach sufficient amplitude, non-axisymmetric structures grow, forming azimuthal vortices in the regions of lowered $\omega_z$. This phenomenon can possibly be attributed to the Rossby Wave Instability, as suggested by the authors. The shear layers exhibit some ripples in the meridional plane (e.g., Fig. 13 in that work), which are possibly produced by the KHI.

For all parameters considered in \cite{Richard2016}, the reduction of $\omega_z$ in between the shear layers with respect to its Keplerian value appears lower than the value $\Delta \omega_z/\Omega = -1/2$ that would result from the formation of uniform-$j_z$ zones (for sufficiently small $\partial_\phi v_R$). This may be a consequence of the non-axisymmetric transport of angular momentum, which should tend to counteract the formation of axisymmetric uniform-$j_z$ regions by the VSI \citep[Paper I and][]{PapaloizouPringle1984}. However, the fact that $\Delta \omega_z <0$ is still indicative that regions of lowered-$\nabla j_z$ are formed (before non-axisymmetric structures are formed, $\omega_z\propto \partial_R j_z$). While we cannot determine from these results whether such regions still maintain some degree of axisymmetry above the midplane, a reduction of $\nabla j_z$ would mean that $\tau_b$ should overcome $\tau_c$ in those locations. It is then possible that a baroclinic amplification of eddies occurs in those simulations, provided this is not restricted by the employed vertical resolution. However, it is challenging to identify this phenomenon in the data presented in that work (Fig. 13) without resorting to 3D vortex detection methods. It is also unclear whether coherent regions of reduced-$\nabla j_z$ can in some cases be destroyed by non-axisymmetric modes, as it may be the case near the midplane in Fig. 8 of \cite{Richard2016}.

Despite the mentioned numerical challenges, the long-term evolution of the VSI and the eventual baroclinic amplification of meridional eddies are problems worth studying in the future. Firstly, besides potentially intervening in the VSI saturation, amplified eddies may have an impact on dust evolution by enhancing the stirring of small dust grains, increasing their average collisional velocities, and even concentrating them in cases where $\Omega_v$ does not exceed $\sim\Omega$ \citep{KlahrHenning1997}, for instance during their acceleration stage. Furthermore, understanding the formation and evolution of meridional vortices may cast some light on the birth of azimuthal anticyclonic vortices produced in 3D VSI simulations \citep{Richard2016,Manger2018,Flock2020,Pfeil2021} and potentially observed in images of protoplanetary disks \citep[e.g.,][]{vanderMarel,deBoer2021,Marr2022}, which can boost planet formation by acting as dust traps. If the KHI has a role in the initial formation of such vortices \citep{LatterPapaloizou2018}, it is likely that these are affected by baroclinic torques in their early stages. Moreover, it is also possible that some vortices form as a result of the breakup of reduced-$\nabla j_z$ bands by non-axisymmetric modes.



\subsection{Stability regions and observational consequences}\label{SS:ObsConsequences}

The fact that our disk models are VSI-stable close to the midplane from a minimum radius is consistent with current dust continuum observations of protoplanetary disks according to the argument by some authors \citep{Flock2017RadHydro,Dullemond2022} that a suppression of the VSI, at least in the outer disk, is required to explain the thin distributions of millimeter-sized dust seen in ALMA dust continuum images of protoplanetary disks, for instance in HD 163296 at $r\approx 100$ au \citep{Doi2021} and in Oph163131 up to $\sim 170$ au \citep{Villenave2022} \citep[see also][]{Pinte2016}. In particular, a suppression of the VSI at large radii could explain why the dust scale height estimated in \cite{Doi2021} is close to $H$ at $68$ au but much smaller at $100$ au. The stable regions in the outer parts of our disk models grow in size as the density of small grains is reduced, which is why it has been proposed that these can naturally occur as a consequence of small-dust depletion due to coagulation \citep{Fukuhara2021,Dullemond2022,Pfeil2023}. To this picture, our local analysis adds the possibility that the VSI may still operate in surface layers even at distances from the star where it is suppressed at the midplane.


Considering the works by \cite{Stoll2016} and \cite{Flock2017RadHydro}, we expect a VSI activity confined to regions above the irradiation surface to produce significant vertical stirring of small dust grains, and therefore to limit vertical settling in such regions. We conjecture that this phenomenon can contribute to the vertically extended appearance of edge-on disks at optical and near-infrared wavelengths \citep{Villenave2020} even at radii where turbulence appears weak or nonexistent at the disk midplane \citep{Villenave2022}, as we expect the vertical stirring of dust grains to push the location of the $\tau=1$ surface for photon scattering at $\mu$m wavelengths away from the midplane. This scenario can be tested in numerical experiments by computing the distributions of different dust populations in a dust fluid approach \citep[e.g.,][]{Krapp2022} and using these to produce synthetic dust continuum and scattered light images. It must be noted, however, that the distributions of small grains in the disk upper layers may also depend on nonideal MHD effects and magnetized winds \citep{Riols2018,Booth2021,Hutchison2021}.

Depending on whether the $\tau=1$ surfaces of $^{12}$CO lines are located inside the unstable surface layers or not (if these occur), it might be possible to detect signs of velocity structures produced by the VSI in such layers in ALMA CO kinematics observations, as proposed by \cite{BarrazaAlfaro2021}. In the same work, it is concluded that the nonthermal spectral broadening of molecular lines produced by the VSI should be negligible compared to current ALMA capabilities, which is consistent with current measured upper bounds \citep[e.g.,][]{Teague2018,Flaherty2020}, and which also applies to an instability localized at surface layers.

\subsection{Additional physics affecting the disk stability}\label{SS:AdditionalEffects}

Some physical effects have been left out of our stability analysis. On the one hand, the disk middle layer can be stabilized for large enough vertically integrated dust-to-gas mass ratios ($\Sigma_\mathrm{dust}/\Sigma_\mathrm{gas} \gtrsim q H/R$) due to the additional buoyancy caused by dust back-reaction \citep{Lin2019}.
On the other hand, the strength of the VSI at the surface layers should depend on the large-scale magnetic field and on the degree of ionization in such locations, which are in general poorly constrained \citep[][]{Lesur2022PPVII}. In particular, recent studies on the interplay of the VSI and magnetic phenomena indicate that strong magnetic fields well coupled to the gas in highly ionized regions (i.e., in the ideal MHD limit) should suppress the VSI by providing additional buoyancy via magnetic tension \citep[][]{LatterPapaloizou2018,CuiLin2021}, whereas, in poorly ionized regions, nonideal MHD effects such as the Hall effect and ambipolar and Ohmic diffusion can diminish this effect and favor the growth of the VSI \citep{CuiBai2020,CuiLin2021,LatterKunz2022,CuiBai2022}. Most of these models assume locally isothermal disks and therefore favor VSI growth. 
Future nonideal MHD studies including finite thermal relaxation times or radiative transfer, ideally self-consistently estimating the local ionization degree \citep[e.g.,][]{Delage2022}, are therefore required in order to better constrain the expected degree of turbulence at the surface layers of protoplanetary disks.

%% file: conclusions.tex
\section{Conclusions}\label{S:Conclusions}

In this work we studied the growth and evolution of secondary instabilities parasitic to the VSI modes and their relation with the VSI saturation in the axisymmetric Rad-HD simulations presented in Paper I. We also extended the VSI-stability analysis of Paper I including the effects of dust-gas thermal equilibration and gas molecular emission, and applied it to construct stability maps in our disk models. Our main findings can be summarized as follows:

\begin{enumerate}
\setlength\itemsep{1em}
    \item In its linear stage, the VSI produces adjacent upward- and downward-directed flows which are initially disconnected. The transport of angular momentum across constant-$j_z$ surfaces reduces the angular momentum gradient between flows moving away from the midplane and the immediate outward flows moving toward the midplane, causing the baroclinic torque ($\tau_b$) to prevail over the centrifugal torque ($\tau_c$) in those regions. Such adjacent flows eventually become connected in upper disk regions, either due to the angular momentum excess of outward-moving gas parcels or due to baroclinic acceleration.
    \item Connected adjacent vertical flows exchange angular momentum in such a way that bands of approximately uniform $j_z$ are formed after a few orbits. The initial $\tau_c$ is consequently suppressed inside of these bands, while the initial $\tau_b$ is largely unaffected, leading to the production of vorticity in those regions by baroclinic torquing. This may explain why the saturated VSI-induced flows survive despite the reduced angular momentum gradient inside of the bands. 
    \item While the meridional shear is also reduced inside of the formed uniform-$j_z$ bands, it is maximum at the interface between them, with large enough values to overcome the stabilizing effect of the radial $j_z$ gradient. As a result, the KHI is triggered at those layers. This instability transfers significant kinetic energy from the VSI flows to KH eddies, acting as an effective friction between the vertical flows and limiting the maximum kinetic energy these can gain due to the VSI driving, possibly leading to the saturation of the VSI.
    \item Large resolutions favor the KHI by increasing the shear in between uniform-$j_z$ bands and reducing the numerical diffusivity. Together with the increase of the number of KH-unstable interfaces with resolution, this may explain why both the average kinetic energy and the Reynolds stresses after saturation decrease with resolution without converging in our resolution range. We require about $100$ cells per scale height to unequivocally identify the KH eddies in vorticity distributions, but the velocity perturbations in the VSI modes show that this instability is triggered in our lower resolution runs as well. Similar signs of developing KHI in other works suggest that this is a rather general phenomenon involved in the VSI saturation. 
    \item Due to the flow pattern produced by the VSI, the initially produced KH eddies rotate in the direction disfavored by $\tau_b$, namely, with $\sgn{\omega_\phi}=-\sgn{\tau_b}$. 
    However, the nonlinear evolution of the KHI at the shear layers also generates counter-rotating eddies in adjacent zones with $\sgn{\omega_\phi}=\sgn{\tau_b}$. As a result, some of those vortices get accelerated by $\tau_b$ in the uniform-$j_z$ zones, possibly with some contribution of vorticity advection from regions near the shear layers. Vortices accelerated in this way reach Mach numbers up to $\sim 0.4$ for $f_\mathrm{dg}=10^{-3}$ and $\sim 0.2$ for $f_\mathrm{dg}=10^{-4}$, corresponding in each case to the maximum velocities in the simulation domain. Conversely, the growth of vortices with $\sgn{\omega_\phi}=-\sgn{\tau_b}$ is disfavored by $\tau_b$. 
    Besides \cite{Klahr2023b}, this mechanism has not been seen in other works due to the high resolutions required to resolve the amplified vortices ($\gtrsim 100$ cells per scale height with our integration methods). 
    \item Our stability analysis shows that the infrequent dust-gas collisions at the disk surface layers can locally suppress the VSI, but these regions can still be unstable for reasonably high gas emissivity.
    \item 
    An extension of this stability analysis to regions in our disk models left out of the simulation domain shows that, for sufficient depletion of small grains via dust coagulation and sufficiently high molecular emissivity, protoplanetary disks can be VSI-unstable in surface layers up to several tens of au while remaining stable at the midplane. This picture is consistent with current observations of vertically extended edge-on disks at optical and near-infrared wavelengths \citep[e.g.,][]{Villenave2020} appearing as thin layers in ALMA dust continuum images \citep[e.g.,][]{Villenave2022}, evidencing a low level of turbulent dust stirring at the midplane. The stability of the surface layers is subject to the gas molecular composition and their local temperature. 
\end{enumerate}

If a baroclinic amplification of meridional eddies does indeed occur in protoplanetary disks, it may affect dust coagulation and fragmentation by increasing the collisional velocities of small dust grains, affecting their diffusivity, and potentially concentrating them \citep{KlahrHenning1997}. On top of this, this process could play a role in the birth of large-scale azimuthal vortices. It is however unclear whether bands of constant specific angular momentum can form in real disks, as these are unstable to non-axisymmetric modes which cannot be captured by our axisymmetric simulations. However, reduced-$\nabla j_z$ bands seem to be present in a number of 3D simulations \citep[][]{Richard2016,Manger2018,Flock2020}. Since the formation of such regions is a requirement for the formation of amplified eddies, determining whether this process can occur and, in that case, how it may interact with non-axisymmetric modes requires rather expensive high-resolution 3D simulations. These would make it possible to explore the
long-term evolution of the VSI and its induced secondary instabilities without running into artifacts produced in axisymmetric simulations, such as the seemingly unrestricted growth of large uniform-$j_z$ regions.

Further work is also needed to make better predictions on the stability of disk surface layers. In particular, improvements to this work can be made by modeling the distribution and temperature of small dust grains, which determine the collisional timescale and therefore affect the strength of the VSI. Conversely, the balance between vertical stirring and settling of dust grains depends on the level of turbulence, and therefore a self-consistent model is needed. Such models should also consider the likely VSI suppression in strongly ionized regions, and, conversely, the expected prevalence of the VSI in regions where nonideal MHD effects become important \citep[e.g.,][]{LatterKunz2022,CuiBai2022}.

%% file: trel.tex
\section{Vorticity generation}\label{A:VorticityGeneration}

In this appendix we derive a series of expressions determining the generation and transport of vorticity in axisymmetric disks. We start by considering a closed streamline $C$ in the $(R,z)$-plane, defined such that its tangent vector is at all locations parallel to $\mathbf{v}_p=\mathbf{v}-(\mathbf{v} \cdot \bm{\hat{\phi}})\bm{\hat{\phi}}=v_R \bm{\hat{R}}+v_z \bm{\hat{z}}$, this is, the projection of $\mathbf{v}$ onto the $(R,z)$-plane, where $\{\bm{\hat{R}},\bm{\hat{\phi}},\bm{\hat{z}}\}$ are the cylindrical unit vectors. By defining the circulation of $\mathbf{v}$ along this curve as the line integral
\begin{equation}
    \Gamma_C(t)=\oint_{C} \mathbf{v}\cdot d\bm{\ell}\,,
\end{equation}
and taking into account the variation of $C$ with time, specifically, $\frac{d\bm{\ell}}{dt}=\mathbf{v}_p(\bm{\ell})$, we can follow analogous steps as in the common derivation of Kelvin's circulation theorem \citep[e.g.,][]{KunduCohen} to obtain
\begin{equation}
\begin{split}
    \frac{d \Gamma_{C}}{dt} &= 
    \oint_{C}\left(\partial_t \mathbf{v}_p 
    + \mathbf{v}_p\cdot \nabla\mathbf{v}_p \right) \cdot d\bm{\ell}\\
    &= 
    \oint_{C} \left(\partial_t \mathbf{v} 
    + \mathbf{v}\cdot \nabla\mathbf{v} + 
    R \Omega^2 \bm{\hat{R}}
    \right) \cdot d\bm{\ell}
\end{split}
\end{equation}
where we have used for the second step the expression of $\nabla\mathbf{v}_p$ in cylindrical coordinates for axisymmetric flows and the fact that $\bm{\hat{\phi}} \cdot d\bm{\ell}
=0$. Replacing the velocity terms using the Navier-Stokes momentum equation, $\partial_t \mathbf{v} 
    + \mathbf{v}\cdot \nabla\mathbf{v}=-\frac{\nabla p}{\rho}-\nabla\Phi$\,,
and using Stokes' theorem to convert the right-hand side of this equation into a surface integral, we arrive to the desired expression,
\begin{equation}\label{Eq:dtVortLineIntegral_ms}
    \frac{d \Gamma_{C}}{dt} = 
    \int_{S} \left(
    \kappa_z^2\, \bm{\hat{\phi}} +
    \frac{\nabla \rho \times \nabla p}{\rho^2}
    \right) \cdot d\bm{S}\,,
\end{equation}
where $C=\partial S$.
The left-hand side of this equation can also be transformed using Stokes' theorem, yielding the following equation for the vorticity ($\bm{\omega}=\nabla\times \mathbf{v}$):
\begin{equation}\label{Eq:dtVortSurfIntegral}
    \frac{d}{dt} \int_{S} \bm{\omega} \cdot d\bm{S} = 
    \int_{S} \left(
    \kappa_z^2\, \bm{\hat{\phi}} +
    \frac{\nabla \rho \times \nabla p}{\rho^2}
    \right) \cdot d\bm{S}\,.
\end{equation}
Finally, taking $d\mathbf{S}=dS\,\hat{\bm{\phi}}$ (i.e., choosing clockwise circulation for the computation of $\Gamma_C$), Eqs. \eqref{Eq:dtVortLineIntegral_ms} and \eqref{Eq:dtVortSurfIntegral} become
Eqs. \eqref{Eq:Circulation} and \eqref{Eq:VorticityIntegral}.

An equation quantifying the local evolution of the $\phi$ component of the vorticity can also be obtained by taking the curl of the momentum equation, leading to
\begin{equation}
    \partial_t \bm{\omega} +
    \nabla\times\left(\mathbf{v}\cdot\nabla\mathbf{v}\right)=
    \frac{\nabla \rho \times \nabla p}{\rho^2} \,.
\end{equation}
The left-hand side of this expression can be simplified by making use of the identity $\mathbf{v}\cdot\nabla\mathbf{v}=\bm{\omega}\times\mathbf{v}+\frac{1}{2}\nabla v^2$\,, after which the resulting term $\nabla\times\left(\mathbf{v}\cdot\nabla\mathbf{v}\right)=\nabla\times\left(\bm{\omega}\times\mathbf{v}\right)$ can be converted into $\nabla\cdot(\bm{\omega}\mathbf{v})-\bm{\omega}\cdot\nabla\mathbf{v}$, resulting in
\begin{equation}
    \partial_t \bm{\omega} +
    \nabla\cdot\left(\bm{\omega}\mathbf{v}\right) =
    \frac{\nabla \rho \times \nabla p}{\rho^2}+ \bm{\omega}\cdot\nabla \mathbf{v}\,.
\end{equation}
This expression is valid in general for 3D flows, as we have not yet assumed axisymmetry. Here it is noteworthy that, unlike in strictly 2D flows, in axisymmetric flows the vortex stretching term $\bm{\omega}\cdot\nabla\mathbf{v}$ is not necessarily zero provided $v_\phi\neq 0$. Equation \eqref{Eq:Vorticity} is finally obtained as the $\phi$ component of this equation after zeroing all $\phi$ derivatives.


\section{Spectral analysis}\label{A:Fourier}

In Section \ref{SS:SpectralAnalysis}, we show energy spectra obtained by operating on Fourier-transformed velocity and momentum distributions. Since computing discrete Fourier transforms requires employing an evenly spaced grid, we take a small portion of the $(r,\theta)$ domain of approximately equal $r$- and $z$-extents, $L$, and rescale it by converting it into a uniform grid defined in a new domain $\{(x_1,x_2)\in [0,L]\times [0,L]\}$. We do so by defining each distribution $V$ in the new grid as $V(x_1(i),x_2(j))=V(r(i),\theta(N_2-j))$, with $i= 1,\ldots,N_1$ and $j= 1,\ldots,N_2$, where $N_1\times N_2$ is the grid resolution of the chosen domain region. To enforce the periodicity of the transformed distributions, we then perform a fourfold replication of our data on an extended grid defined in the domain $S=\{(x_1,x_2)\in [-L,L]\times [-L,L]\}$ with resolution $2 N_1\times 2 N_2$. We achieve this by assigning each considered distribution $V$ to a single quadrant of the new grid and reflecting it across the $x_1$ and $x_2$ axes.

To compute energy spectra, we follow a procedure similar to that in \cite{Alexakis2018} with some modifications that allow us to consider nonuniform $\rho$ distributions \citep[a similar approach is followed, for example, in][]{Sengupta2023}. We define the Fourier transform of a periodic function $F$ defined on $S$ as
\begin{equation}
    \widetilde{F}(\mathbf{k})=
    \left\langle F(\mathbf{r}) e^{-i \mathbf{k} \cdot \mathbf{x}} \right\rangle\,,
\end{equation}
where $\mathbf{k}$ is the wavenumber and $\langle\cdot\rangle$ denotes average over $S$, in such a way that
\begin{equation}
    F(\mathbf{x})=\sum_\mathbf{k}
    \widetilde{F}(\mathbf{k})e^{i \mathbf{k} \cdot \mathbf{x}}\,.
\end{equation}
With these definitions, it is straightforward to show that the volume average of the meridional kinetic energy, $\epsilon_p=\frac{\mathbf{m}_p\cdot \mathbf{v}_p}{2}$, can be written as
\begin{equation}
    \langle \epsilon_p \rangle
    =\frac{1}{2}
    \sum_\mathbf{k} \widetilde{\mathbf{m}}^*_{p}(\mathbf{k})
    \cdot \widetilde{\mathbf{v}}_{p}(\mathbf{k})\,,
\end{equation}
where $^*$ denotes complex conjugation, while $\mathbf{m}_p=\rho \mathbf{v}_p$ denotes the momentum density in the $(R,z)$-plane (see Appendix \ref{A:VorticityGeneration}). Defining a wavenumber norm step $\Delta k$, we can then use this expression to approximate the kinetic energy spectrum as 
\begin{equation}
    E(k) = \frac{1}{2\Delta k}
    \sum_{k \leq ||\mathbf{k}|| < k+ \Delta k}
    \widetilde{\mathbf{m}}^*_{p}(\mathbf{k})
    \cdot \widetilde{\mathbf{v}}_{p}(\mathbf{k})\,,
\end{equation}
in such a way that $E(k)\Delta k$ is the kinetic energy within a shell in $\mathbf{k}$-space of radius $k$ and width $\Delta k$.

We take the $\mathbf{v}_p$ and $\mathbf{m}_p$ distributions from the region $r\in[5.25,5.75]$ au and $\theta \in[\theta_\mathrm{min}+\delta,\theta_\mathrm{min}+\delta+\Delta \theta]$, where $\Delta \theta$ is such that $L=0.5$ au, which is approximately equal to the local pressure scale height $H=c_s \Omega^{-1}$ at that location. We choose this upper layer region since it exhibits VSI growth for both considered dust-to-gas ratios, and we choose $\delta=0.02$ to consider a region separate from the $\theta$-boundary. The domain is large enough that it contains between $5$ and $6$ radial VSI wavelengths, and small enough that the maximum relative stretching of cells when replacing the logarithmic $r$-grid with the uniform $x1$-grid is below $5\%$. Since in this domain $\rho$ decreases with $z$ down to a factor of $1/20$, the obtained spectra are mostly representative of the kinetic energy distribution in the largest-density region.

\section{Thermal relaxation timescale}\label{A:CollisionalTimescale}

\subsection{Dust-gas collisional timescales}

In general, the relaxation timescale $t_\mathrm{rel}$ for gas temperature perturbations differs from $t_\mathrm{cool}$ if the timescale of energy transfer between dust and gas particles is long enough that the dust and gas temperatures, $T_d$ and $T_g$ respectively, can be different during the relaxation. To consider this effect, we obtain expressions for $t_\mathrm{rel}$ following an analogous procedure to the analysis in \cite{Barranco2018}. As in Paper I (Appendix B), we neglect advection by taking $\mathbf{v}=\mathbf{0}$. We also neglect gas opacities and assume the heating rate of spherical dust grains due to collisions with gas particles to be given, as in \cite{Burke1983}, by
\begin{equation}
    \Lambda_{dg}(a)=2 \mathcal{A}_H\,\pi a^2 \,
    n_g \overline{v_g}
    k_B(T_g-T_d)\,,
\end{equation}
where $a$ is the grain radius, $n_g=\rho/(\mu u)$ and $\overline{v_g}=\sqrt{\frac{8 k_B T}{\pi \mu u}}$ are the number density and the mean thermal speed of gas particles, respectively, while $\mathcal{A}_H$ is the thermal accommodation coefficient for atomic and molecular hydrogen interacting with graphite and silicate grains, which we henceforth assume to be $\sim 0.5$ \citep{Burke1983}.

Let us first assume that all dust grains are at the same temperature. Using the grain size distribution $n(a)\sim a^{-3.5}$ assumed in our employed opacity model, we can compute the total energy exchange rate per unit volume between gas and dust as
\begin{equation}
    \Lambda_{dg}=
    I[ \pi a^2 ]\,
    n_g \overline{v_g}
    k_B(T_g-T_d)\,,
\end{equation}
with $I[f(a)]=\int_{a_\mathrm{min}}^{a_\mathrm{max}}da\,f(a) n(a)$, where $a_\mathrm{min}$ and $a_\mathrm{max}$ are respectively the minimum and maximum grain effective radii. The  size distribution is normalized in such a way that the total dust density $\rho^\mathrm{tot}_d$
(i.e., not the density of small grains $\rho_d$ used for the computation of opacities)
can be computed as $I\left[\frac{4\pi a^3}{3}\rho_{gr}\right]$, where $\rho_{gr}$ is the bulk density of the dust grains, and thus $I[\pi a^2]=\frac{3 \rho^\mathrm{tot}_d}{4\rho_{gr}}(a_\mathrm{max}a_\mathrm{min})^{-1/2}$. The resulting equations for the evolution of the temperature of gas and dust particles are
\begin{equation}\label{Eq:CoolEqGasDust}
    \begin{split}
        c_g \rho \, \partial_t T_g &= -\Lambda_{dg} \\
        c_d \rho^\mathrm{tot}_d \, \partial_t  T_d &= \Lambda_{dg}
        + c\,G^0 -\nabla \cdot \mathbf{F}_\mathrm{Irr}\,,
    \end{split}
\end{equation}
where $c_g=\frac{k_B}{(\Gamma-1)\mu u}$ and $c_d$ are the gas and dust specific heat capacities at constant volume and constant pressure, respectively. The radiative cooling term $G^0$ in this expression is computed by replacing the term $a_R T^4$ (Equation 5 in Paper I) with $a_R T_d^4$. A linearization of these equations for small departures $\delta T_g$ and $\delta T_d$ with respect to the equilibrium temperature $T=T_g=T_d$
leads to the following system:
\begin{equation}\label{Eq:linearizedTdTg}
    \begin{split}
    \partial_t \delta T_g &= -(\delta T_g-\delta T_d)/t_g \\
    \partial_t \delta T_d &= -(\delta  T_d-\delta T_g)/t_d - \delta T_d/t_r
    \,,
    \end{split}
\end{equation}
where $t_g=\frac{4}{3(\Gamma-1)}\left(\frac{\rho_{gr}}{\rho^\mathrm{tot}_d}\right)\frac{(a_\mathrm{max}a_\mathrm{min})^{1/2}}{\overline{v_g}}$, $t_d=t_g\left(\frac{\rho^{tot}_d}{\rho}\right)\left(\frac{c_d}{c_g}\right)$, and $t_r=\left(\frac{\rho_d^\mathrm{tot}}{\rho_d}\right)\frac{c_d T}{c\kappa^d_P [4a_RT^4+b_P(a_RT^4-E_r)]}$, with $b_P=\frac{d \log \kappa_P^d}{d\log T}$. In these equations we have dropped the $\delta E_r$ term arising from the $cG^0$ term, since the present analysis is only required in optically thin regions (more generally, for optically thin perturbations) where $\delta E_r\ll \delta (a_R T^4)$ (see Appendix B in Paper I). The gas thermal relaxation time can then be obtained in terms of the evolution of $\delta T_g$ given an initial condition of the form $(\delta T_g,\delta T_d)=(\delta T_0,0)$. In that case, the solution of Equation \eqref{Eq:linearizedTdTg} is well approximated by $\delta T_g = \delta T_0\, e^{-t/t_\mathrm{rel}}$, with
\begin{equation}
    t_\mathrm{rel}
    = 2 t_{||}\left[1-\sqrt{
    1-4t_{||}^2/(t_gt_r)
    }\right]^{-1}\,,
\end{equation}
where $t_{||}^{-1}=t_g^{-1}+t_d^{-1}+t_r^{-1}$. As shown in \cite{Barranco2018}, an approximation to this expression can be obtained 
under the condition that $t_g\gg t_d,t_r$, in which case $t_\mathrm{rel}$ is well approximated by
\begin{equation}\label{Eq:t_rel0}
    t_\mathrm{rel}
    \approx t_g(1+t_r/t_d) =
    t_g+t_\mathrm{cool,thin}\,,
\end{equation}
where $t_\mathrm{cool,thin}$ is the optically thin limit of the radiative cooling time (Equations 16 and 17 in Paper I). This approximation was based on the disk parameters in that work, where grain sizes in the range $[1,100]$ cm were considered, whereas in the present work we assume the much smaller $a_\mathrm{max}=19$ $\mu$m and $1.8$ mm for $f_\mathrm{dg}=10^{-3}$ and $10^{-4}$, respectively. 
This is a more reasonable assumption in the regions where collisional times become relevant, where we can expect centimeter-sized grains to be removed by vertical settling, as shown for example in the model by \cite{Flock2020}. Yet, with our employed size distribution, the relations $t_r,t_d\ll t_g$ are still satisfied, the latter of which we have verified taking the values of the specific heats capacities of graphite and amorphous silicates from \cite{Butland1973} and \cite{ZellerPohl1971}, respectively. Therefore, as we have verified in our disk models, the approximation in Eq. \eqref{Eq:t_rel0} is still valid with $0.1\%$ and $0.01\%$ accuracy for $f_\mathrm{dg}=10^{-3}$ and $f_\mathrm{dg}=10^{-4}$, respectively.  Moreover, since in optically thick regions $t_g\ll t_\mathrm{cool}$, we can replace Equation \eqref{Eq:t_rel0} with the following expression, valid in all radiative transport regimes:
\begin{equation}\label{Eq:t_rel}
    t_\mathrm{rel}
    \approx t_g+t_\mathrm{cool}\,,
\end{equation}
where $t_\mathrm{cool}$ is computed as in Paper I (Equation 16). The resulting collisional timescale, $t_g$, is independent on the gas density, as both the heat capacity $\rho c_g$ and the energy exchange rate $\Lambda_{dg}$ in Equation \eqref{Eq:CoolEqGasDust} are proportional to $\rho$.

Away from the midplane, dust settling might impose a lower cutoff to the dust distribution than the assumed $a_\mathrm{max}$ satisfying $\rho_d^\mathrm{tot}/\rho=10^{-2}$, especially in the dust-depleted case, in which $a_\mathrm{max}=1.8$ mm. However, this should not affect the value of $t_g$, which depends on $a_\mathrm{max}$ as $t_g\propto \sqrt{a_\mathrm{max}a_\mathrm{min}}/\rho_d^\mathrm{tot}$. Since $\rho_d^\mathrm{tot}\propto \sqrt{a_\mathrm{max}}-\sqrt{a_\mathrm{min}}$ and $a_\mathrm{min}\ll a_\mathrm{max}$, this results in $t_g\propto \sqrt{a_\mathrm{max}a_\mathrm{min}}/(\sqrt{a_\mathrm{max}}-\sqrt{a_\mathrm{min}})\approx \sqrt{a_\mathrm{min}}$, and thus $t_\mathrm{g}$ is rather insensitive to the precise value of $a_\mathrm{max}$. 


\subsection{Inclusion of gas radiative emission}

The same analysis can be applied if the emission of radiation by gas molecules is taken into account. We assume for this that the gas emits thermally at a rate determined by the Planck-averaged gas absorption opacity $\kappa^g_P(T)$ (Kirchhoff's law), and so we include a gas-radiation interaction terms of the same form of $G^0$, specifically $G^{0}_g=\kappa^g_P(T)\rho(E_r-a_R T^4)$. This leads to the following system of equations:
\begin{equation}
    \begin{split}
        c_g \rho \, \partial_t T_g &= -\Lambda_{dg} + c\,G^0_g\\
        c_d \rho^\mathrm{tot}_d \, \partial_t  T_d &= \Lambda_{dg}
        + c\,G^0 -\nabla \cdot \mathbf{F}_\mathrm{Irr}\,,
    \end{split}
\end{equation}
which for small perturbations with respect to equilibrium becomes
\begin{equation}\label{Eq:linearizedTdTg_trg}
    \begin{split}
    \partial_t \delta T_g &= -(\delta T_g-\delta T_d)/t_g - \delta T_g/t_{r,g} \\
    \partial_t \delta T_d &= -(\delta  T_d-\delta T_g)/t_d - \delta T_d/t_r
    \,,
    \end{split}
\end{equation}
where $t_{r,g}=\frac{c_g T}{c \kappa_P^g [4a_RT^4+b^g_P(a_RT^4-E_r)]}$, with $b^g_P=\frac{d \log \kappa_P^g}{d\log T}$.
In deriving Equation \eqref{Eq:linearizedTdTg_trg} we have again assumed $\delta E_r \ll \delta (a_R T^4)$, which is valid for optically thin perturbations. A solution of this system for initial $T_g$ perturbations leads to the optically thin thermal relaxation timescale,
\begin{equation}
    t_\mathrm{rel,thin}
    = 2 t_{||}\left[1-\sqrt{
    1-4t_{||}^2(t_g^{-1}t_r^{-1}+t_{r,g}^{-1}(t_d^{-1}+t_r^{-1}))}\right]^{-1}\,,
\end{equation}
where now $t_{||}^{-1}=t_g^{-1}+t_d^{-1}+t_r^{-1}+t^{-1}_{r,g}$. This expression can be further approximated by noting that the factor proportional to $t^2_{||}$ is much smaller than $1$ for low enough $t^{-1}_{r,g}$, namely, for low enough $\kappa^g_P$, and that $t_g t_r/t_{r,g}\ll 1$ for our disk parameters, which leads to
\begin{equation}\label{Eq:trelkPg}
    t_\mathrm{rel,thin}\approx 
    \left(
    \frac{1}{t_g+t_\mathrm{cool,thin}}
    +
    \frac{1}{t_{r,g}}
    \right)^{-1}\,. 
\end{equation}
The lowest accuracy of this approximation in our disk models, achieved for our highest considered value of $\kappa_P^g=1$ cm$^{2}$ g$^{-1}$, is $0.1\%$ and $0.01\%$ for $f_\mathrm{dg}=10^{-3}$ and $f_\mathrm{dg}=10^{-4}$, respectively. For largely different $t_g$, $t_{r,g}$, and $t_\mathrm{cool,thin}$, Equation \eqref{Eq:trelkPg} is equivalent to
\begin{equation}
t_\mathrm{rel,thin}\approx\min(\max(t_\mathrm{cool,thin},t_g),t_{r,g})\,,
\end{equation}
which is the limit of the relaxation timescale considered in \cite{Malygin2017} for vanishing gas-gas collisional times (albeit with some differences in the way of computing $t_g$).

In dense regions where $t_g\ll t_\mathrm{cool,thin}$, we can safely assume $T_g=T_d$ and denote these temperatures by $T$. In that case, $t_\mathrm{rel}$ tends to the radiative cooling time that can be obtained by repeating the linear analysis in Paper I (Appendix B) including gas opacities. This can be achieved by simply replacing the  Planck and Rosseland photon mean free paths due to dust absorption by the corresponding combined mean free paths for dust and gas absorption, namely,
\begin{equation}
\begin{split}
    \lambda_P &= \frac{1}{\kappa_P\rho_d } \rightarrow \lambda^{dg}_P
    = \frac{1}{\kappa^{dg}_P\rho}\\
    \lambda_R &= \frac{1}{\kappa_R\rho_d } \rightarrow \lambda^{dg}_R = \frac{1}{\kappa^{dg}_R\rho}\,,
\end{split}
\end{equation}
where $\kappa^{dg}_P= f_\mathrm{dg}\kappa^d_P + \kappa^g_P$ and $\kappa_R^\mathrm{dg}$ are respectively the Planck and Rosseland means of $f_\mathrm{dg}\kappa^d_\nu+\kappa_\nu^g$, while $\kappa^d_\nu$ and $\kappa^g_\nu$ are the frequency-dependent dust and gas absorption opacities per dust and gas mass, respectively (scattering is neglected in this work, but otherwise $\kappa^{dg}_R$ should be replaced by the total opacity coefficient $\chi^{dg}_R$). In this way, the same system of equations as in Paper I (Appendix B) is obtained, and the resulting cooling time is
\begin{equation}
    t_\mathrm{cool} = t_\mathrm{cool,thin} 
    \frac{3+\lambda^{dg}_R\lambda^{dg}_P k^2}{\lambda^{dg}_R\lambda^{dg}_P k^2}\,,
\end{equation}
where
\begin{equation}
    t_\mathrm{cool,thin}= \frac{\rho \epsilon}{4\, a_R T^4+ b^{dg}_P(a_R T^4-E_r)}\frac{\lambda^{dg}_P}{c}
\end{equation}
is the optically thin limit of this expression valid for $\lambda^{dg}_R\lambda^{dg}_P k^2\gg 1$, with $b^{dg}_P=\frac{d \log \kappa^{dg}_P}{d \log T}$. With this result, we can write a general expression for the relaxation timescale valid in all radiative transport and collisional regimes, given by
\begin{equation}\label{Eq:trel_appendix}
    t_\mathrm{rel} = t_\mathrm{rel,thin}  \frac{3+\lambda^{dg}_R\lambda^{dg}_P k^2}{\lambda^{dg}_R\lambda^{dg}_P k^2}\,,
\end{equation}
where $t_\mathrm{rel,thin}$ is given by Equation \eqref{Eq:trelkPg}. Equation \eqref{Eq:trel_appendix} also comprises the cases in which dust-gas collisional times and gas opacities are not included. In the optically thick limit, in which case $t_g \ll t_\mathrm{cool}$, this expression tends to the optically thick limit of $t_\mathrm{cool}$. For optically thin perturbations for which $t_g \ll t_\mathrm{cool}$, it tends to $t_\mathrm{cool,thin}$. Finally, in regions where collisional times become important, the gas perturbations are always optically thin, and thus this expression tends to the correct limit $t_\mathrm{rel,thin}$.

\subsection{Size-dependent dust temperature}

We lastly explore how the thermal relaxation timescale is changed if the dust temperature has a dependence on size $T_d(a)\propto a^{-b}$ arising from the frequency dependence of the grain emissivity (\cite{Glassgold2004}, where $b\in[1/5,1/3]$, and \cite{Pauly2016}, where $b=1/6$), in which case the collisional energy exchange rate per unit volume becomes
\begin{equation}
    \Lambda_{dg}=
    I[ \pi a^2 ]\,
    n_g \overline{v_g}
    k_B(T_g-B\langle T_d\rangle)\,,
\end{equation}
where $\langle T_d\rangle=I[T_d(a)]/I[1]$ is the average dust temperature and 
\begin{equation}
\begin{split}
B &= \frac{1+2 b/5}{1+2b}
\frac{1-(a_\mathrm{min}/a_\mathrm{max})^{5/2}}{1-(a_\mathrm{min}/a_\mathrm{max})^{5/2+b}}
\frac{1-(a_\mathrm{min}/a_\mathrm{max})^{1/2+b}}{1-(a_\mathrm{min}/a_\mathrm{max})^{1/2}}\\
&\approx \frac{1+2 b/5}{1+2b}\lesssim 1\,.
\end{split}
\end{equation}
Taking $\kappa^g_P=0$, this leads to
\begin{equation}
    \begin{split}
        c_g \rho \, \partial_t T_g &= -\Lambda_{dg} \\
        \overline{c_d} \rho^\mathrm{tot}_d \, \partial_t \langle T_d\rangle &= \Lambda_{dg}
        + c\,G^0 -\nabla \cdot \mathbf{F}_\mathrm{Irr}\,,
    \end{split}
\end{equation}
where $\overline{c_d}\approx \frac{1+2b/5}{1-2b}c_d$ for $b<1/2$. The radiative cooling term $cG^0$ is now computed using the dust average temperature replacing $T_d^4$ with $\langle T_d^4 \rangle\approx \frac{(1+2b/5)^4}{(1+8b/5)}\langle T_d \rangle^4\approx\langle T_d \rangle^4$, where in the last term we dropped $\mathcal{O}(b^2)$ terms. Thus, linearizing these equations for small temperature perturbations, we obtain the system
\begin{equation}
    \begin{split}
    \partial_t \delta T_g &= -(\delta T_g-B\delta \langle T_d\rangle)/t_g \\
    \partial_t \delta \langle T_d\rangle &= -(B\delta  \langle T_d\rangle-\delta T_g)/t_d - \delta \langle T_d\rangle/t_r
    \,,
    \end{split}
\end{equation}
where now $t_d=t_g(\rho^\mathrm{tot}_d/\rho)(\overline{c_d}/c_g)$, and $t_r=\overline{c_d}/(4\kappa_P a_RT^3)$. The resulting optically thin thermal relaxation time is
\begin{equation}
    t_\mathrm{rel,thin}
    = 2 t_{||}\left[1-\sqrt{
    1-\frac{4t_{||}^2}{t_g}\left(
    \frac{1}{t_r}
    +\frac{1-B^2}{t_d}
    \right)
    }\right]^{-1}\,,
\end{equation}
which is well approximated by
\begin{equation}\label{Eq:t_rel_B}
    t_\mathrm{rel,thin} \approx \frac{t_g+t_\mathrm{cool,thin}}{1+(t_\mathrm{cool,thin}/t_g)(1-B^2)}
    \,,
\end{equation}
given that both $t^2_{||}/(t_g t_r)$ and $t^2_{||}/(t_g t_d)$ are $\ll 1$. Since the size dependence of $T_d$ results from the balance between absorption of starlight and infrared emission, values of $B \neq 1$ can only be assumed above the irradiation surface, where $t_\mathrm{cool,thin}<t_g$. Therefore, Equation \eqref{Eq:t_rel_B} shows that a slow dependence of the dust temperature with the grain size does not significantly alter the thermal relaxation time. The same conclusion is reached if gas radiative emission is included, in which case the optically thin relaxation time is given by Equation \eqref{Eq:trelkPg} replacing $t_g+t_\mathrm{cool,thin}$ with the right-hand side of Equation \eqref{Eq:t_rel_B}.